\documentclass[10pt,journal,compsoc]{IEEEtran}
\usepackage{amsmath,amsfonts}
\usepackage{algorithmic}
\usepackage{algorithm}
\usepackage{array}
\usepackage[caption=false,font=normalsize,labelfont=sf,textfont=sf]{subfig}
\usepackage{textcomp}
\usepackage{stfloats}
\usepackage{url}
\usepackage{verbatim}
\usepackage{graphicx}
\usepackage{cite}

\usepackage{tikz}
\usepackage{amsmath}

\usepackage{hyperref}

\usepackage{amssymb}

\usepackage{fancybox}
\usepackage{fbox}
\usepackage{comment}
\usepackage{bm}

\usepackage{multirow}

\usepackage{array}

\usepackage{pifont}

\newcommand{\crossmark}{\ding{55}}

\begin{document}

\title{Almost-Free Queue Jumping for Prior Inputs in Private Neural Inference}

\author{
Qiao~Zhang,
        Minghui~Xu,Tingchuang~Zhang,
        and~Xiuzhen~Cheng,~\IEEEmembership{Fellow,~IEEE}
\thanks{Q. Zhang, M. Xu, T. Zhang , and X. Cheng are with the School of Computer Science and Technology, Shandong University, Qingdao 266237, China (e-mail:qiao.zhang@sdu.edu.cn; mhxu@sdu.edu.cn; tczhang@mail.sdu.edu.cn; xzcheng@sdu.edu.cn).}}


\IEEEpubid{\begin{minipage}{\textwidth}\ \\[12pt]
This work has been submitted to the IEEE for possible publication. Copyright may be transferred without notice, after which this version may no longer be accessible.
\end{minipage}}
\maketitle
\begin{abstract}
Privacy-Preserving Machine Learning as a Service (PP-MLaaS) enables secure neural network inference by integrating cryptographic primitives such as homomorphic encryption (HE) and multi-party computation (MPC), protecting both client data and server models. Recent mixed-primitive frameworks have significantly improved inference efficiency, yet they process batched inputs sequentially, offering little flexibility for prioritizing urgent requests. Naïve queue jumping introduces considerable computational and communication overhead, increasing non-negligible latency for in-queue inputs.

We initiate the study of privacy-preserving queue jumping in batched inference and propose PrivQJ, a novel framework that enables efficient priority handling without degrading overall system performance. PrivQJ exploits shared computation across inputs via in-processing slot recycling, allowing prior inputs to be piggybacked onto ongoing batch computation with almost no additional cryptographic cost. Both theoretical analysis and experimental results demonstrate over an order-of-magnitude reduction in overhead compared to state-of-the-art PP-MLaaS systems.
\end{abstract}

\begin{IEEEkeywords}
Machine Learning as a Service, Neural Model, Private Inference, Homomorphic Encryption, Multi-Party Computation.
\end{IEEEkeywords}

\section{Introduction}
\IEEEPARstart{A}{s} Machine Learning (ML) models become widely adopted in numerous applications and model training is always data-hungry and hardware-demanding, the end users or clients opt to send their inputs to a resource-abundant server that possesses high-performance models to obtain accurate prediction in a cost-efficient manner. However, concerns for data privacy make it challenging to deploy such a service in practice and the privacy concerns mainly lie in two folds: (1) The clients need to send input to the cloud server but they prefer to prevent any other parties including cloud server from obtaining her private input and model output, and there are also regulations to prohibit clients from disclosing data such as the GDPR~\cite{zaeem2020effect} for business information and the HIPPA~\cite{ness2007influence} for medical records, and (2) The server is reluctant to make its model parameters public since they could be intellectual property~\cite{cryptoeprint:2025/1365}.

Privacy-Preserving Machine Learning as a Service (PP-MLaaS) aims to achieve data protection throughout the interaction between client and server. 
Specifically, as an ML model contains a stack of linear and non-linear functions, PP-MLaaS intends to specifically integrate crypto primitives, e.g., Multi-Party Computation (MPC)~\cite{demmler2015aby} and Homomorphic 
\IEEEpubidadjcol Encryption (HE)~\cite{fan2012somewhat}, to sequentially compute each function of the neural model, with encrypted input from the client and proprietary model parameters from the server. Such integration guarantees that (1) the client finally receives the model
output with its input blind to the server while (2) the server’s model
parameters remain hidden to the client beyond model output. While PP-MLaaS provides a promising way to align MLaaS with data protection, the key challenge is how to efficiently integrate cryptographic primitives into neural network computations. Without addressing this, there could be extremely high computational complexity or lower prediction accuracy, resulting from either large circuit sizes or function approximations.

To make PP-MLaaS more practical, a series of elegant works have been proposed to accelerate such process~\cite{cryptonets,liu2017oblivious,juvekar2018gazelle,mishra2020delphi,rathee2020cryptflow2,zhang2021gala,hussain2021coinn,huang2022cheetah,qiaosp24,pangsp24,cryptoeprint:2024/136}.
Remarkably, there has been a dramatic enhancement in inference efficiency, with exponential improvements from early frameworks like CryptoNets~\cite{cryptonets} to more advanced systems such as CrypTFlow2~\cite{rathee2020cryptflow2} and Cheetah~\cite{huang2022cheetah}. At a conceptual level, these privacy-preserving frameworks bolster computational efficiency by strategically optimizing appropriate cryptographic techniques to execute linear functions (e.g., dot product and convolution) and nonlinear functions (e.g., ReLU) within neural networks. Major cryptographic primitives that are frequently utilized in these frameworks include HE~\cite{brakerski2012fully,fan2012somewhat,cheon2017homomorphic}, and MPC techniques like Oblivious Transfer (OT)~\cite{brassard1986all}, Secret Sharing (SS)~\cite{shamir1979share}, and Garbled Circuits (GC)~\cite{bellare2012foundations,yao1986generate}. Most notably, the
mixed-primitive frameworks that leverage HE for linear calculations while integrating MPC for nonlinear operations have been particularly effective in enhancing efficiency of PP-MLaaS~\cite{liu2017oblivious,juvekar2018gazelle,rathee2020cryptflow2,huang2022cheetah,qiaosp24}, and this paper advances the optimization of efficiency in mixed-primitive mode.

\begin{figure}[!tbp]
\centering
\includegraphics[trim={6.42cm 0.9cm 0.05cm 0.3cm}, clip, scale=0.423]{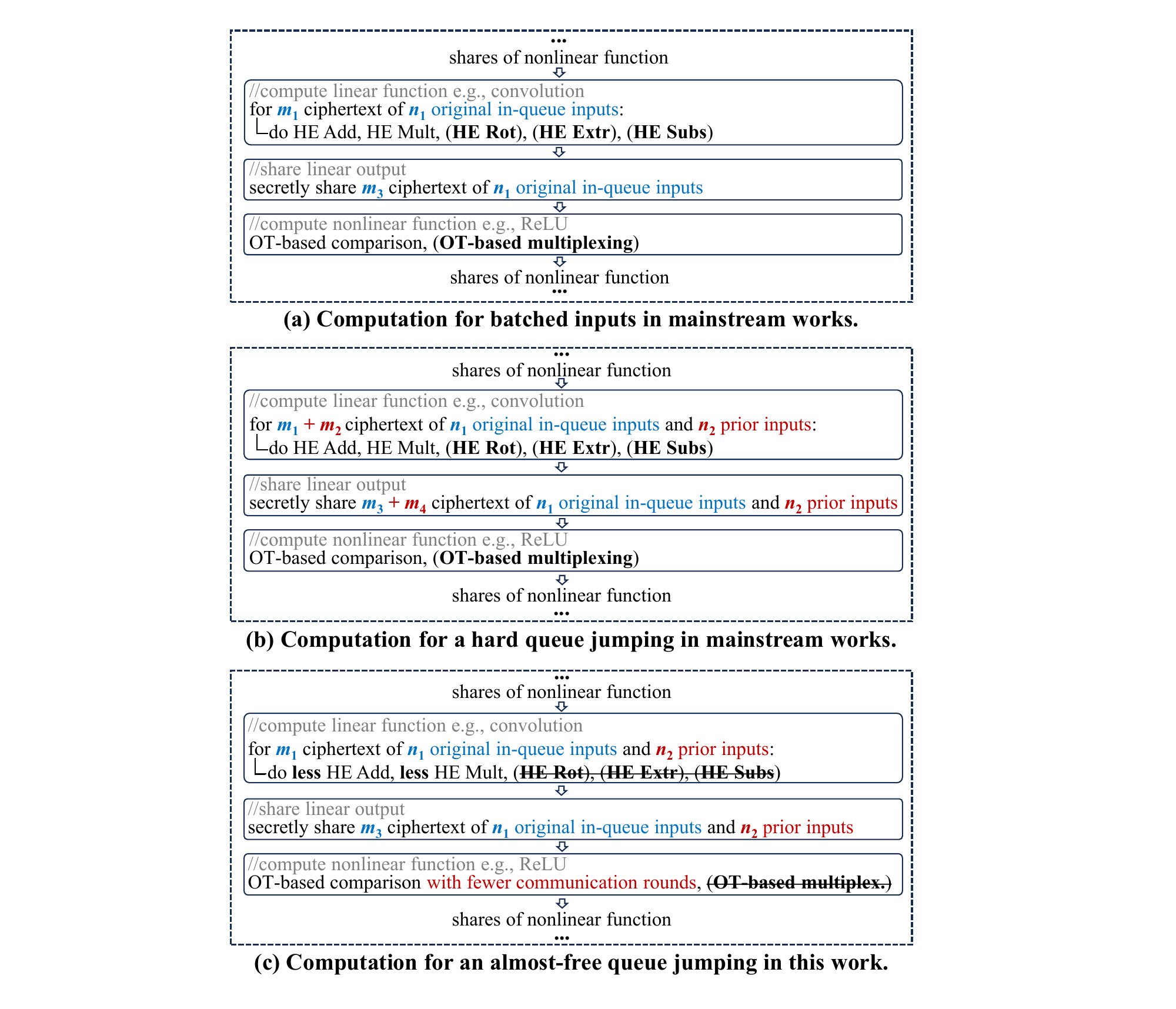}
\caption{Comparison of processing logic for batched inputs.}
\label{fig:compare}
\end{figure}
\textbf{State-Of-The-Art.} In mainstream PP-MLaaS frameworks including mixed-primitive ones, the same procedure is performed for different inputs to compute each function in a neural model. Given a queue of private inputs at the client, a common way to complete PP-MLaaS for the whole inputs is to form a set of input batches and all inputs in each batch are first encrypted and sent from the client to the server, and then both parties involve in computing the function output and obtaining their respective output shares of the first function in neural model. Since one neural model is composed of a stack of linear and nonlinear functions, the above output shares at both parties serve as the input for the second function, and the process of input share feeding, share-based computation, and output share generation is sequentially applied to every function until the client finally obtains model outputs for all inputs in one batch. In this way, the PP-MLaaS for all inputs in a queue is performed batch by batch as illustrated in Figure~\ref{fig:compare}(a), and
the batch is either in individual style where the ciphertext for each input is independent from that of others~\cite{rathee2020cryptflow2,huang2022cheetah,qiaosp24,mishra2020delphi,liu2017oblivious,juvekar2018gazelle} or in group style where one ciphertext could be associated with multiple inputs~\cite{cryptoeprint:2024/136,cryptonets}.

On the other hand, the newly-coming inputs in the queue could be prior or urgent such that waiting in order is undesired or even intolerable. For example, an emergent patient in a local hospital's diagnosis list needs to get the output in advance from the central hospital that has a high-performance medical neural model. In such case, queue jumping is unavoidable. When it happens, there involves a hard queue jumping to foremost batch as shown in Figure~\ref{fig:compare}(b) and the waiting cost added for original in-queue inputs which now follow after the queue-jumping ones is either proportional to the complexity of computing individual input~\cite{rathee2020cryptflow2,huang2022cheetah,qiaosp24} if the number of inputs in foremost batch is temporarily increased to include queue-jumping inputs, or equivalent to the overhead of computing an entire batch~\cite{cryptonets,cryptoeprint:2024/136} if the batch size is fixed and at least one original in-queue input in foremost batch needs to be dropped out to the subsequent batch. In both cases, the added waiting cost is non-negligible, ranging from hours to even days~\cite{qiaosp24,cryptoeprint:2024/136}, which provides limited flexibility to the private inference system in case of emergency and degraded efficiency to those affected in-queue inputs.

\begin{table*}[!t]
\centering
\scriptsize
\caption{Waiting cost added to original in-queue inputs which now follow after queue-jumping inputs namely the prior ones. The complexity comparison is for computing adjacent nonlinear ReLU function and linear convolution (Conv) function, which are the main functions in standard neural networks. Conv takes as input a 3-dimension matrix with size $C_i\times{H_i}\times{W_i}$. $C_o$ and $f_h$ are the number and the size of filters, respectively. $H_o$ and $W_o$ are the height and width of each 2-dimension convolution output. The output of ReLU acts as the input of Conv, with input data $\bm{x}$ in size $C_i\times{H_i}\times{W_i}$. $\mathsf{rd}$ is the number of involved communication rounds. $f_{\mathsf{r}}'(\bm{x})$ and {Mx}$(\bm{x})$ are the derivative of ReLU for $\bm{x}$ and the multiplexing to obtain ReLU output based on shares of $f_{\mathsf{r}}'(\bm{x})$ and $\bm{x}$, respectively. The communication cost in all schemes excludes common overhead for computing $f_{\mathsf{r}}'(\bm{x})$.}
\label{complexity:overview}
\resizebox{\textwidth}{!}{%
\begin{tabular}{ c | c c c c | c c }
\hline\hline
 \multirow{2}{*}{Schemes} & \multicolumn{4}{c|}{Computation  Cost for HE Operations} & \multicolumn{2}{c}{Communication Cost} \\
\cline{2-7}
 & \# Rot & \# Extr & \# Mult (\# Add) & \# Dec& \# Ciphertexts & \# Run-time Round\\
\hline\hline
 CryptoNets\cite{cryptonets}$^*$ & 0 & -&$O(C_iC_oH_oW_o(f_h)^2)$& $O(C_oH_oW_o)$& $O(C_iH_iW_i+C_oH_oW_o)$& $1+\textsf{rd}_{\{f_{\textsf{r}}'(\bm{x}), \textsf{Mx}(\bm{x})\}}$\\
 
 MiniONN\cite{liu2017oblivious} & 0 & -&$O(C_iC_oH_oW_o(f_h)^2)$& \begin{tabular}{@{}c@{}}$O(C_iC_oH_oW_o$\\$(f_h)^2/N)$\end{tabular}& \begin{tabular}{@{}c@{}}$O((C_iC_oH_oW_o$\\$(f_h)^2)/N)$\end{tabular}& $0.5+\textsf{rd}_{\{f_{\textsf{r}}'(\bm{x}), \textsf{Mx}(\bm{x})\}}$\\
 
 GAZELLE\cite{juvekar2018gazelle} & \begin{tabular}{@{}c@{}}$O(C_iH_iW_i(f_h)^2$\\$+C_iC_oH_iW_i)$\end{tabular} & -&$O(C_iC_oH_iW_i(f_h)^2)$& $O(C_oH_oW_o/N)$& $O((C_iH_iW_i+C_oH_oW_o)/N)$& $1+\textsf{rd}_{\{f_{\textsf{r}}'(\bm{x}), \textsf{Mx}(\bm{x})\}}$\\
 
 DELPHI\cite{mishra2020delphi} & \begin{tabular}{@{}c@{}}$O(C_iH_iW_i(f_h)^2$\\$+C_iC_oH_iW_i)$\end{tabular} & -&$O(C_iC_oH_iW_i(f_h)^2)$& $O(C_oH_oW_o/N)$& $O((C_iH_iW_i+C_oH_oW_o)/N)$& $0.5+\textsf{rd}_{\{f_{\textsf{r}}'(\bm{x}), \textsf{Mx}(\bm{x})\}}$\\
 
 CrypTFlow2\cite{rathee2020cryptflow2} & \begin{tabular}{@{}c@{}}$O(C_iH_iW_i(f_h)^2$\\$+C_o)$\end{tabular} & -&$O(C_iC_oH_iW_i(f_h)^2)$& $O(C_oH_oW_o/N)$& $O((C_iH_iW_i+C_oH_oW_o)/N)$& $1+\textsf{rd}_{\{f_{\textsf{r}}'(\bm{x}), \textsf{Mx}(\bm{x})\}}$\\
 
 Cheetah\cite{huang2022cheetah} &0  & $O(C_oH_oW_o)$ & $O(C_iC_oH_iW_i)$ & $O(C_oH_oW_o)$& \begin{tabular}{@{}c@{}}$O(C_iH_iW_i/N$\\$+C_oH_oW_o)$\end{tabular}&$1+\textsf{rd}_{\{f_{\textsf{r}}'(\bm{x}),\textsf{Mx}(\bm{x})\}}$\\
 
 {FIT}\cite{qiaosp24} & 0& 0& $O(C_iC_oH_oW_o(f_h)^2)$& \begin{tabular}{@{}c@{}}$O(C_iH_iW_i/N$\\$+C_o)$\end{tabular}& \begin{tabular}{@{}c@{}}$O(C_iC_oH_oW_o$\\$(f_h)^2/N+C_o)$\end{tabular} & $1+\textsf{rd}_{\{f_{\textsf{r}}'(\bm{x})\}}$\\
 
 NEXUS\cite{cryptoeprint:2024/136}$^*$ &0 &0 &$O(C_iC_o(f_h)^2)$ &$O(C_o)$ & \begin{tabular}{@{}c@{}}$O(C_iC_o(f_h)^2/N$\\$+C_o)$\end{tabular} &$1+\textsf{rd}_{\{f_{\textsf{r}}'(\bm{x}), \textsf{Mx}(\bm{x})\}}$\\
 
 \textbf{PrivQJ} (this work)&0 &0 &0 & 0&0 &$0.5$\\
\hline\hline
\multicolumn{7}{l}{$^*$ Tailored for mixed-primitive inference.}
\end{tabular}
}
\end{table*}
\textbf{Key Observations.} Table~\ref{complexity:overview} further demonstrates concrete complexity added to original in-queue inputs that are forced to follow after the prior inputs when applying a hard queue jumping based on mainstream works for batched-input inference. Specifically, the main overhead originates from additional HE computation, such as massive HE Addition (Add), Multiplication (Mult), Extraction (Extr), and expensive HE Rotation (Rot) for linear function, as well as the extra ciphertext needed to carry computed result for prior inputs, and we observe that the invocation of such non-trivial computation and communication is due to \textit{the repeated processing logic for different inputs}, which has to form totally new ciphertext to perform corresponding computation for new input like the prior ones. Different from the relatively independent treatment to each input, we explore the possibility to jointly utilize shared computation of all inputs in a small batch, with the goal of enabling more efficient queue jumping that does not hurt the efficiency for original in-queue inputs, especially those that follow after the prior ones.

We begin the exploration by seeking any computation and communication redundancy towards the inputs in a batch, and the data layout in ciphertext for each input is therefore investigated since a ciphertext includes a number of slots to confidentially carry equal amount of plaintext values and any slots that do not directly involve in the computation for corresponding in-queue input might be utilized to enable the computation for prior input in a Single-Instruction-Multiple-Data (SIMD) manner. However, given the data packing process in mainstream works, it is challenging to  make effective utilization of any peripheral slots from one input to simultaneously perform computation for another input. This is because the number of slots that are available for meaningful HE operations, after data packing process based on mainstream works, is either zero \cite{cryptonets,huang2022cheetah,liu2017oblivious} or lower than the minimum that is needed to do similar computation for prior input, e.g., one channel \cite{rathee2020cryptflow2,mishra2020delphi,juvekar2018gazelle} or at least an entire input \cite{qiaosp24, cryptoeprint:2024/136}.  

\textbf{Overall Methodology.} While it seems impossible to break such limitation to make use of those available slots, we shift to view the queue jumping problem in batched-input inference as a jigsaw puzzle where computation for each in-queue input is able to output a disjoint and unique piece of result for prior input and the joint computation among in-queue inputs in a size-optimal batch correctly produces the complete result for prior inputs. In this way, computation for in-queue inputs in the batch is able to establish compatible correlation with prior inputs, which provides us opportunity to perform a more efficient queue jumping through SIMD. As such, we propose PrivQJ, a privacy-preserving queue jumping framework that features with an in-processing slot recycling among in-queue inputs in a batch, which limits the introduction of any crypto cost needed for prior inputs. As shown in Figure~\ref{fig:compare}(c), PrivQJ is capable of piggybacking the computation for prior input within the computing process for minimum in-queue inputs, which achieves an almost-free computation for prior input without any expensive HE Rot or additional HE enablers such as HE Extr and HE Substitution (Subs) \cite{cryptoeprint:2024/136}, and with fewer communication rounds. In a nutshell, the \textit{contributions} of this work are summarized as follows.
\begin{itemize}
\item We initiate the problem of privacy-preserving queue jumping in batched-input inference and propose our solution called PrivQJ, which tackles non-trivial waiting cost added to the in-queue inputs.

\item PrivQJ is able to fully utilize the shared computation ability within a batch, which largely reduces the extra waiting cost as shown in Table~\ref{complexity:overview}. Notably, PrivQJ eliminates any HE cost needed for prior inputs.

\item Quantitative complexity justifies the efficiency advantages of PrivQJ and experimental results further confirm over 10 times cost reduction with moderate batch size, and the source code is available \href{https://anonymous.4open.science/r/PrivQJ-13C2}{\textit{here}}.
\end{itemize}

\textbf{Road Map}. The rest of the paper is organized as follows. In Section~\ref{sec:preliminary}, we introduce threat model, neural network, and the crypto primitives that are adopted in this work. Section~\ref{sys:description} describes the concrete design of PrivQJ, and analyzes the corresponding complexity and security. The evaluation results are elaborated in Section~\ref{eval}. Section \ref{related:work} gives an overview of PP-MLaaS frameworks and Section~\ref{conclusion} concludes the paper.

\section{Preliminaries}
\label{sec:preliminary}

This section first describes the threat model of \textrm{PrivQJ} and then introduces the basic components in the neural model. Finally we present the adopted crypto primitives.


For ease of understanding, we highlight the \textit{notations} that are commonly used in the rest of this paper: Given $u_1,u_2\in\mathbb{R}$ and $j\in\mathbb{N}$,
$[j]$ is denoted as the integer set $\{1, \dots, j\}$ while $[u_1,u_2]=\{u|u_1\leq u\leq u_2\}$.
$\lceil\cdot\rceil$ and $\left \lfloor\cdot\right \rfloor$ are the ceiling function and flooring function, respectively. The random sampling of an element $r$ from a distribution $\mathcal{D}$ is denoted as $r$ $\scriptstyle{\gets}{\$}$ $\mathcal{D}$. Given a prime $p\in\mathbb{N}$, $\mathbb{Z}_{p}=\mathbb{Z}\cap[-\lfloor p/2\rfloor,\lfloor p/2\rfloor]$ for $p>2$ while $\mathbb{Z}_{2}=\{0,1\}$. Meanwhile, $+, -,$ and $\times$ are element-wise addition, element-wise subtraction, and element-wise multiplication, respectively, in either ciphertext domain or plaintext domain depending on whether any ciphertext is involved or not. {Meanwhile, we denote \fbox{$\bm{a}$} and \doublebox{$\bm{a}$} as the \textit{$\mathcal{S}$-encrypted HE ciphertext} and \textit{$\mathcal{C}$-encrypted HE ciphertext} whose corresponding decryption return the plaintext vector $\bm{a}\in\mathbb{Z}_{p}^N$ where $N$ is the number of slots in one ciphertext}. Furthermore, all values thereafter are respectively scaled to $\mathbb{Z}_{p}$ such that the computation is performed within prime moduli $p$.


\subsection{Threat Model}
This work considers a classic 2-party privacy-preserving MLaaS scenario where the client $\mathcal{C}$ sends the encrypted input to the cloud server $\mathcal{S}$ that has a well-trained neural model. The subsequent computation is conducted from the first function to the last one of the neural model, according to specifically-designed privacy-preserving protocol until $\mathcal{C}$ finally obtains the model output, with it's input blind to $\mathcal{S}$ and $\mathcal{S}$'s model parameters hidden to $\mathcal{C}$.

Meanwhile, \textrm{PrivQJ} assumes the semi-honest adversary model that is adopted among state-of-the-art PP-MLaaS frameworks~\cite{liu2017oblivious,rathee2020cryptflow2,huang2022cheetah,qiaosp24,cryptoeprint:2024/136}. Concretely, both $\mathcal{C}$ and $\mathcal{S}$ follow the protocol while each of them tries to obtain extra knowledge based on the received information. We prove in Section~\ref{sys:complexity} that \textrm{PrivQJ} is secure against semi-honest adversaries.

\subsection{Neural Network}
As for the neural model at server, we considers the long-standing Convolutional Neural Networks (CNNs) which have been applied in various data-sensitive scenarios  such as the C-Lung-RADS system for diagnosis of pulmonary nodules~\cite{WOS:001314921600001} and the Sybil framework for prediction of lung cancer~\cite{2023Sybil}. 
Specifically, a CNN contains a stack of linear functions and non-linear functions and an input is sequentially computed from the first function to the last one, and the output of a function serves as the input of next function. Meanwhile, similar computation is applied for a batch of inputs from the first function to the last one.

\textbf{Linear functions}.
\textit{Convolution} $f_{\textrm{c}}(\bm{x})$: Given the kernel $\bm{k}$ in a neural model as $\bm{k}=(\bm{k}_1|\bm{k}_2|\cdots|\bm{k}_{C_o})\in\mathbb{Z}_p^{C_o\times C_i\times H_f\times W_f}$ where $\bm{k}_j\in\mathbb{Z}_p^{C_i\times H_f\times W_f}$ with $j\in[C_o]$ is the $j$-th filters, and an input as $\bm{x}\in\mathbb{Z}_p^{C_i\times H_i\times W_i}$, $f_{\textrm{c}}(\bm{x})$ outputs $C_o$ channels, each of which is with size $H_o$-by-${W_o}$ and is obtained by gradually moving one of $C_o$ filters over $\bm{x}$ with stride $s\in\mathbb{N}^{+}$. Specifically, an element in $j$-th output channel is the sum of all ${C_i}{H_f}W_f$ values in $\bm{k}_j$, each of which is scaled by the value from $\bm{x}$ that locates at the same position within filter window of $\bm{k}_j$, and the next element in $j$-th output channel is obtained by first moving $\bm{k}_j$ over $\bm{x}$ with stride $s$ and then performing similar computation between $\bm{k}_j$ and those values from $\bm{x}$ that are within the filter window. Here $H_o$ and $W_o$ are the height and width of output channel, respectively. $H_i$ (or $H_f$) is the height of each input channel (or each filter) while $W_i$ (or $W_f$) is the corresponding width.

\textit{Dot Product} $f_{\textrm{d}}(\bm{x})$: Given the model weight $\bm{w}$ with size $n_o\times{n_i}$ and an input $\bm{x}$ with length $n_i$, $f_{\textrm{d}}(\bm{x})$ outputs a vector with length $n_o$ where the $j$-th element ($j\in[n_o]$) is the sum of all multiplied values from element-wise multiplication between $\bm{x}$ and $j$-th row of $\bm{w}$. Here $n_i$ and $n_o$ are the length of input vector and the length of output vector, respectively.

\textit{MeanPooling} $f_{\textrm{mn}}(\bm{x})$: Given the input $\bm{x}\in\mathbb{Z}_p^{C_i\times H_i\times W_i}$, $f_{\textrm{mn}}(\bm{x})$ outputs $C_i$ channels each of which is in size ${\left \lceil \frac{H_i}{s_{\textsf{n}}} \right \rceil}\times{\left \lceil \frac{W_i}{s_{\textsf{n}}} \right \rceil}$, and each element in $j$-th output channel ($j\in[C_i]$) is obtained by summing and averaging components from $j$-th input channel $\bm{x}_j\in\mathbb{Z}_p^{H_i\times W_i}$ within the $s_{\textsf{n}}$-by-${s_{\textsf{n}}}$ pooling window where $s_{\textsf{n}}\in\mathbb{N}^+$.

\textit{Batch Normalization} $f_{\textrm{bn}}(\bm{x})$: Given the input $\bm{x}\in\mathbb{Z}_p^{C_i\times H_i\times W_i}$, the output of $f_{\textrm{bn}}(\bm{x})$ is with the same size as $\bm{x}$, and the $j$-th output channel ($j\in[C_i]$) is obtained by first multiplying the $j$-th input channel $\bm{x}_j\in\mathbb{Z}_p^{H_i\times W_i}$ with a scaling factor and then adding a shifting value to that multiplied channel.

\textbf{Non-linear functions}.  
\textit{ReLU} $f_{\textrm{r}}(\bm{x})$: Given input $\bm{x}\in\mathbb{Z}_p^{C_i\times H_i\times W_i}$, the output of $f_{\textrm{r}}(\bm{x})$ is with the same size as $\bm{x}$, and an element in $j$-th output channel ($j\in[C_i]$) turns out to be the maximum between zero and the value from $\bm{x}_j\in\mathbb{Z}_p^{H_i\times W_i}$ at the same position. $f_{\textrm{r}}(\bm{x})$ is alternatively expressed as the element-wise multiplication between $\bm{x}$ and the derivative of ReLU $f'_{\textrm r}(\bm{x})\in\mathbb{Z}_2^{C_i\times H_i\times W_i}$ where an element is 1 if and only if the value from $\bm{x}$ at the same position is positive.

\textit{MaxPooling} $f_{\textrm{mx}}(\bm{x})$: Given the input $\bm{x}\in\mathbb{Z}_p^{C_i\times H_i\times W_i}$, $f_{\textrm{mx}}(\bm{x})$ outputs $C_i$ channels each of which is with size ${\left \lceil \frac{H_i}{s_{\textsf{m}}} \right \rceil}\times{\left \lceil \frac{W_i}{s_{\textsf{m}}} \right \rceil}$, and each element in $j$-th output channel ($j\in[C_i]$) is obtained by selecting the maximum component from $j$-th input channel $\bm{x}_j\in\mathbb{Z}_p^{H_i\times W_i}$ within the $s_{\textsf{m}}$-by-${s_{\textsf{m}}}$ pooling window where $s_{\textsf{m}}\in\mathbb{N}^+$.

\textbf{Remarks}: 
Given a queue of private inputs at $\mathcal{C}$ and the CNN model at $\mathcal{S}$, there could be prior inputs that need to securely obtain model output in advance. A hard queue jumping could lead to non-trivial waiting cost added to the in-line inputs that now follow after the prior ones. As such, an efficient solution for 2-party privacy-preserving neural inference with more resilience to emergent inputs helps to facilitate more stable inference implementation with stringent privacy regulation.

\subsection{Crypto Primitives}\label{pre:crypto}
\textbf{Homomorphic Encryption (HE)}: HE enables linear operations (such as addition and multiplication) over ciphertext without decryption, which is especially suitable for privacy-preserving computation of linear functions~\cite{demmler2015aby}. \textrm{PrivQJ} adopts the widely-used BFV scheme~\cite{fan2012somewhat} as its HE backend to compute the linear functions in neural models. Specifically, given a plaintext vector $\bm{a}\in\mathbb{Z}_{p}^N$, the HE encryption operation transforms $\bm{a}$ into one ciphertext and the HE decryption operation turns the ciphertext back to original plaintext $\bm{a}$. Here one element in $\bm{a}$ is mapped into one slot in the ciphertext and recall that $N$ is the number of slots in one ciphertext. 

We use two HE operations besides the HE encryption and decryption, namely HE Addition (HE Add) and HE constant Multiplication (HE cMult). HE Add enables addition between the HE ciphertext of $\bm{a}$ and the HE ciphertext of $\bm{b}\in\mathbb{Z}_{p}^N$ (or just plaintext $\bm{b}$), and the added ciphertext is the encryption of a vector that is the element-wise sum of $\bm{a}$ and $\bm{b}$. HE cMult performs multiplication between plaintext $\bm{a}$ and the HE ciphertext of $\bm{b}$, and the multiplied ciphertext is the encryption of a vector that is the element-wise multiplication of $\bm{a}$ and $\bm{b}$. {Recall that \fbox{$\bm{a}$} and \doublebox{$\bm{a}$} are respectively the $\mathcal{S}$-\textit{encrypted ciphertext} and $\mathcal{C}$-\textit{encrypted ciphertext} whose corresponding decryption return the plaintext vector $\bm{a}$.}
    
\textbf{Oblivious Transfer (OT)}: In the OT protocol, the sender has multiple messages and the receiver holds selection bits. At the end of the protocol, the receiver obtains the message whose index corresponds to the exact selection bits, without knowing any information about other messages, and the sender does not know which message the receiver has chosen in the meantime.
    
\textbf{Additive Secret Sharing (ASS)}: ASS enables a party to share its secret $x\in\mathbb{Z}_p^*$ by first masking $x$ with randomness $x_0=r{\scriptstyle{\gets}{\$}}$ $\mathbb{Z}_p^*$
and then sending the noised piece $x_1=x-r$ mod $p$ to the other party. The secret is recovered by summing up $x_1$ with the randomness $x_0$ as $x=x_0+x_1$ mod $p$.

\textbf{Remarks}: The PP-MLaaS frameworks with mixed primitives, where HE is adopted to compute linear functions and OT is applied to calculate non-linear functions, have demonstrated particular effectiveness to boost inference efficiency. Given a queue of private inputs, a practical way to finish privacy-preserving inference for them is to perform computation batch by batch. Under such case, it would cause an un-negligible issue that a long waiting time is added to original in-queue inputs which now follow after inserted prior inputs when the queue jumping happens. To tackle this problem, this work enables a full utilization of available slots among all ciphertext of foremost input batch such that all HE computation for prior inputs is piggybacked in the computation for original in-queue inputs in that batch, effectively reducing the added cost within fewer communication rounds in case of emergent queue jumping.

\section{Privacy-Preserving Queue Jumping}\label{sys:description}
\begin{figure}[!tb]
\centering
\includegraphics[trim={2.8cm 5.2cm 11.5cm 0cm}, clip, scale=0.79]{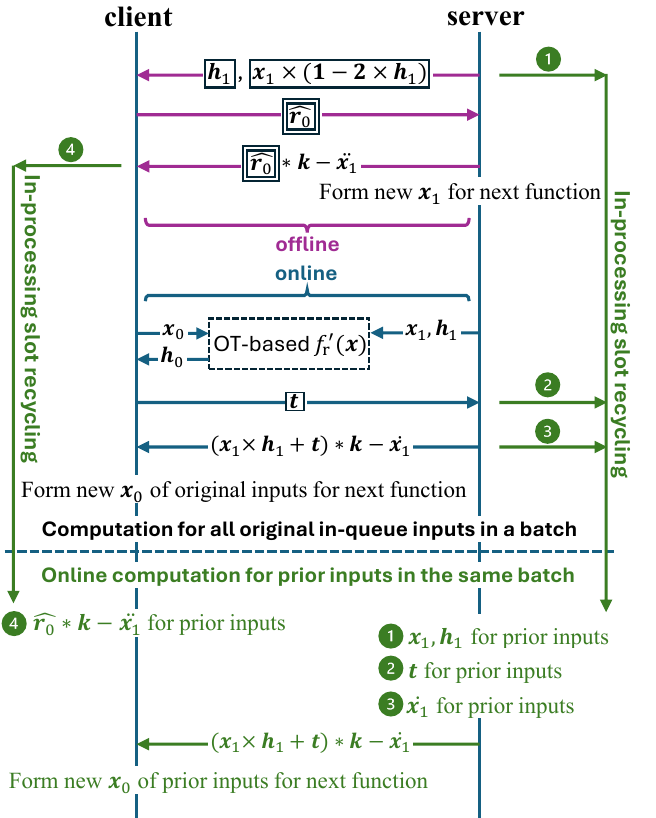}
\caption{Computation for two adjacent functions in a batch.}
\label{fig:reluconv}
\end{figure}
\subsection{Overview of PrivQJ}
Figure \ref{fig:reluconv} illustrates how PrivQJ embeds the computation for high-priority (prior) inputs into the batched secure inference workflow, without disrupting the computation process for original in-queue inputs in the same batch. Generally speaking, PrivQJ features with an in-processing slot recycling strategy within the computation for two adjacent nonlinear-linear functions, without any extra HE cost for prior inputs and with only one half communication round in addition, i.e., 0.5 round.

Specifically, PrivQJ is mainly composed of two parts. The first part is for original in-queue inputs where the computation includes a pre-processing offline and a running-time online. As for the offline, the server $\mathcal{S}$ and the client $\mathcal{C}$ respectively send pre-generated ciphertext to each other to make preparation for the online, and the online involves an OT-based partial computation for non-linear function and a one-round computation for linear function. Such working flow for two adjacent functions puts most computation load to input-independent offline, eliminates steps that are indispensable to obtain output of individual function, reduces the overall offline-online cost, and makes the input-dependent online more efficient compared with other works that optimize each single function. To reduce the non-trivial waiting cost that is added by the hard queue jumping for prior inputs, instead of directly discarding the peripheral slots~\cite{huang2022cheetah,liu2017oblivious,rathee2020cryptflow2,mishra2020delphi,juvekar2018gazelle,qiaosp24, cryptoeprint:2024/136} or having to additionally introduce considerable cipherext to compute for original in-queue inputs that are dropped-out to a new batch~\cite{cryptonets}, PrivQJ shifts to view any available slots in intermediate ciphertext, at both offline and online, as unique pieces of jigsaw puzzle and devises a compact slot recycling computation within already created ciphertext for the batch such that the system derives all intermediate values for the prior inputs with minimal additional operations, e.g., invoking no extra HE operations.

After the online computation for original in-queue inputs in the batch, PrivQJ simultaneously finishes the preparation to compute for prior inputs and a uni-directional communication from $\mathcal{S}$ to $\mathcal{C}$ is followed to form shares for the next function. This slot-recycling mechanism enables PrivQJ to achieve online queue jumping at near-zero extra cost while preserving the integrity and efficiency of the original batched computation.

In the following, the offline-online computation for original in-queue inputs in the batch is first introduced, which is tailored from~\cite{qiaosp24} with less computation cost. Then, concrete mechanism to recycle all slots in ciphertext of original in-queue inputs and the way to determine minimal batch size are described in detail. {For ease of understanding, we recall that \fbox{$\bm{a}$} and \doublebox{$\bm{a}$} are respectively the $\mathcal{S}$-\textit{encrypted ciphertext} and $\mathcal{C}$-\textit{encrypted ciphertext} whose corresponding decryption return the plaintext vector $\bm{a}$.}

\subsection{Tailored Computation for In-queue Inputs}
\label{sys:mechanism}

We start with the computation for adjacent ReLU-Convolution functions $f_{\mathrm c}f_{\mathrm r}(\bm{x})$, which is the main function block in standard neural networks such as AlexNet~\cite{krizhevsky2012imagenet}, VGG~\cite{simonyan2014very} and ResNet~\cite{he2016deep}.
Specifically,
\begin{align}
&f_{\mathrm c}f_{\mathrm r}(\bm{x})={f_{\mathrm r}}(\bm{x})\ast\bm{k}=\{f'_{\mathrm r}(\bm{x})\times \bm{x}\}\ast\bm{k}\;\;\;\mathrm{mod}\;p\\
&=\{(\bm{h}_0+\bm{h}_1-2\times\bm{h}_0\times\bm{h}_1)\times(\bm{x}_0+\bm{x}_1)\}\ast\bm{k}\;\;\;\mathrm{mod}\;p\\
&=\{\bm{x}_0\times\bm{h}_0-\bm{r}_0
+\bm{x}_1\times\bm{h}_1+\underbrace{\bm{x}_0\times(\bm{1}-\bm{2}\times\bm{h}_0)\times\bm{h}_1}_{\mbox{\textcircled{\scriptsize1}}}+
\notag\\
&\quad\;\;\underbrace{\bm{x}_1\times(\bm{1}-\bm{2}\times\bm{h}_1)\times\bm{h}_0}_{\mbox{\textcircled{\scriptsize2}}}\}\ast\bm{k}+\bm{r}_0\ast\bm{k}\;\;\;\mathrm{mod}\;p\label{eq:unfold}
\end{align}
where
\begin{equation}
\left\{
\begin{aligned}
&\bm{h}_0,\bm{h}_1\in\mathbb{Z}_2^{C_i\times H_i\times W_i}\;\mathrm{and}\; \bm{h}_0\oplus\bm{h}_1=f'_{\mathrm r}(\bm{x}),\oplus:\mathrm{XOR}\\
&\bm{r}_0,\bm{x}_0,\bm{x}_1\in\mathbb{Z}_p^{C_i\times H_i\times W_i}\;\mathrm{and}\; \bm{x}_0+\bm{x}_1\;\mathrm{mod}\;p=\bm{x}\\
&\forall{x}\in\mathbb{Z}_p,\;\textrm{we have}\; f'_{\mathrm r}({x})\;\mathrm{is}\;x\;\mathrm{if}\;{x}>0\;\mathrm{and\;else\;is}\;0\\
&\bm{k}\in\mathbb{Z}_p^{C_o\times C_i\times H_f\times W_f}
\end{aligned}
\right.\notag
\end{equation}

Here $f_{\mathrm c}f_{\mathrm r}(\bm{x})$ is decomposed into the sum of two convolutions and the kernel $\bm{k}$ is held by the server, and \textit{the variables with subscript 1 and subscript 0 are the corresponding shares from server and from client, respectively}. Meanwhile, the client-generated randomness $\bm{r}_0$ is introduced in Eq.~\eqref{eq:unfold} such that the server is able to compute the first convolution in plaintext without exposing any information. More concretely, given the capability of $\mathcal{S}$ to pre-generate $\bm{h}_1$ and $\bm{x}_1$, the first convolution in Eq.~\eqref{eq:unfold} is realized by the following two steps: 

Step 1: The client computes three terms at online according to Eq.~\eqref{eq:unfold} as: (1) plaintext $(\bm{x}_0\times\bm{h}_0-\bm{r}_0)$ based on freshly-obtained $\bm{x}_0$ and $\bm{h}_0$, (2) encrypted term {\normalsize{\textcircled{\scriptsize1}}} given available $\mathcal{S}$-encrypted ciphertext \fbox{$\bm{h}_1$}, and (3) encrypted term {\normalsize{\textcircled{\scriptsize2}}} given available $\mathcal{S}$-encrypted ciphertext \fbox{$\bm{x}_1\times(\bm{1}-\bm{2}\times\bm{h}_1)$};

Step 2: The client adds up above three terms and sends the summed ciphertext \fbox{$\bm{t}$} to the server as
\begin{align}
\fbox{$\bm{t}$}&=\{\bm{x}_0\times\bm{h}_0-\bm{r}_0+\underbrace{\bm{x}_0\times(\bm{1}-\bm{2}\times\bm{h}_0)\times\fbox{$\bm{h}_1$}}_{\mbox{\normalsize{\textcircled{\scriptsize1}}}}+\notag\\
&\;\;\underbrace{\fbox{$\bm{x}_1\times(\bm{1}-\bm{2}\times\bm{h}_1)$}\times\bm{h}_0}_{\mbox{\normalsize{\textcircled{\scriptsize2}}}}\}.\label{eq:unfold:v3}
\end{align}
Upon receiving \fbox{$\bm{t}$}, the server performs ciphertext decryption, adds the decrypted plaintext $\bm{t}$ with $(\bm{x}_1\times\bm{h}_1)$, and obtains the first convolution in Eq.~\eqref{eq:unfold}, which we name as 
\[\bm{y}=\{(\bm{x}_1\times\bm{h}_1)+\bm{t}\}\ast\bm{k}.\]

In a nutshell, Eq.~\eqref{eq:unfold} now turns into
\begin{align}
&\{\bm{x}_0\times\bm{h}_0-\bm{r}_0
+\bm{x}_1\times\bm{h}_1+\underbrace{\bm{x}_0\times(\bm{1}-\bm{2}\times\bm{h}_0)\times\fbox{$\bm{h}_1$}}_{\mbox{\textcircled{\scriptsize1}}}+
\notag\\
&\underbrace{\fbox{$\bm{x}_1\times(\bm{1}-\bm{2}\times\bm{h}_1)$}\times\bm{h}_0}_{\mbox{\normalsize{\textcircled{\scriptsize2}}}}\}\ast\bm{k}+\bm{r}_0\ast\bm{k}\label{eq:unfold:v1}\\
&=\{(\bm{x}_1\times\bm{h}_1)+\fbox{$\bm{t}$}\}\ast\bm{k}+\bm{r}_0\ast\bm{k}
\end{align}
and $\mathcal{S}$ first sends \fbox{$\bm{h}_1$} and \fbox{$\bm{x}_1\times(\bm{1}-\bm{2}\times\bm{h}_1)$} to the client at offline. As for the online, the client and the server first involve in the OT-based protocol, e.g.,~\cite{rathee2020cryptflow2}, to get their respective shares for $f'_{\mathrm r}(\bm{x})$ namely $\bm{h}_0$ and $\bm{h}_1$. Here $\bm{h}_1$ has been already pre-generated by $\mathcal{S}$. After that, the first convolution in Eq.~\eqref{eq:unfold} is obtained by $\mathcal{S}$ based on aforementioned Step 1 and Step 2.

As for the second convolution in Eq.~\eqref{eq:unfold} namely $\bm{r}_0\ast\bm{k}$, it is completed offline since both $\bm{r}_0$ and $\bm{k}$ are input-independent. 
Specifically, $\mathcal{C}$ first transforms $\bm{r}_0$ in to matrix $\widehat{\bm{r}_0}\in\mathbb{Z}_p^{H_fW_fC_i\times H_oW_o}$ based on the im2col~\cite{jia2014caffe} which turns convolution computation into matrix-matrix multiplication. Then $\mathcal{C}$ encrypts $\widehat{\bm{r}_0}$ into \doublebox{$\widehat{\bm{r}_0}$} and sends it to $\mathcal{S}$. Upon receiving \doublebox{$\widehat{\bm{r}_0}$}, $\mathcal{S}$ is able to compute convolution between \doublebox{$\widehat{\bm{r}_0}$} and $\bm{k}$ via only HE constant Mult (cMult) and HE Add, and more details are described in Section~\ref{r0kslotrec}. After that, $\mathcal{S}$ masks the convolution result with randomness $\bm{\ddot{x}}_1\in\mathbb{Z}_p^{C_o\times H_o\times W_o}$. As such the masked convolution is finally sent back to $\mathcal{C}$ which decrypts the ciphertext and obtains $(\bm{r}_0\ast\bm{k}-\bm{\ddot{x}}_1)$. In this way, $\mathcal{C}$ and $\mathcal{S}$ share $\bm{r}_0\ast\bm{k}$ by $\mathcal{C}$ holding $(\bm{r}_0\ast\bm{k}-\bm{\ddot{x}}_1)$ and $\mathcal{S}$ holding $\bm{\ddot{x}}_1$. 

To summarize the complete process for each in-queue input in the batch, Eq.~\eqref{eq:unfold} now turns into 

\begin{align}
&\{\bm{x}_0\times\bm{h}_0-\bm{r}_0
+\bm{x}_1\times\bm{h}_1+\underbrace{\bm{x}_0\times(\bm{1}-\bm{2}\times\bm{h}_0)\times\fbox{$\bm{h}_1$}}_{\mbox{\normalsize{\textcircled{\scriptsize1}}}}+
\notag\\
&\underbrace{\fbox{$\bm{x}_1\times(\bm{1}-\bm{2}\times\bm{h}_1)$}\times\bm{h}_0}_{\mbox{\normalsize{\textcircled{\scriptsize2}}}}\}\ast\bm{k}+\doublebox{$\widehat{\bm{r}_0}$}\ast\bm{k}\label{eq:unfold:v2}\\
&=\{(\bm{x}_1\times\bm{h}_1)+\fbox{$\bm{t}$}\}\ast\bm{k}+\doublebox{$\widehat{\bm{r}_0}$}\ast\bm{k}\label{eq:fi1}
\end{align}
and there thus involves an offline-online interaction between the client and the server. 

As for offline, the client and the server independently transmit pre-computed ciphertext in Eq.~\eqref{eq:unfold:v2}: First, the server-encrypted ciphertext, \fbox{$\bm{h}_1$} and \fbox{$\bm{x}_1\times(\bm{1}-\bm{2}\times\bm{h}_1)$}, are sent to $\mathcal{C}$; Second, the client-encrypted ciphertext, \doublebox{$\widehat{\bm{r}_0}$}, is sent to $\mathcal{S}$ and $($\doublebox{$\widehat{\bm{r}_0}$}$\ast\bm{k}-\bm{\ddot{x}}_1)$ is then locally computed and sent back to $\mathcal{C}$ to get plaintext share $(\bm{r}_0\ast\bm{k}-\bm{\ddot{x}}_1)$.

As for online, $\mathcal{S}$ and $\mathcal{C}$ first get their respective shares for $f'_{\mathrm r}(\bm{x})$ (namely $\bm{h}_0$ and $\bm{h}_1$) via OT-based protocol such as~\cite{rathee2020cryptflow2}. Then, $\mathcal{S}$ obtains the first convolution in Eq.~\eqref{eq:unfold}, i.e., $\bm{y}$, according to the description for Eq.~\eqref{eq:unfold:v1}. Here, we reduce the number of convolution from two in \cite{qiaosp24}, which are $(\bm{x}_1\times\bm{h}_1)\ast\bm{k}$ at offline and $\bm{t}\ast\bm{k}$ at online, to only one by directly combining plaintext $\bm{t}$ with $(\bm{x}_1\times\bm{h}_1)$ at negligible overhead. After that, $\mathcal{S}$ shares $\bm{y}$ with $\mathcal{C}$ by sending to $\mathcal{C}$ plaintext $(\bm{y}-\bm{\dot{x}}_1)$ while holding $\bm{\dot{x}}_1\in\mathbb{Z}_p^{C_o\times H_o\times W_o}$ for itself. Till now, the client forms its online share of $f_{\mathrm c}f_{\mathrm r}(\bm{x})=\bm{y}+\bm{r}_0\ast\bm{k}$ as
\[
(\bm{r}_0\ast\bm{k}-\bm{\ddot{x}}_1+\bm{y}-\bm{\dot{x}}_1)=f_{\mathrm c}f_{\mathrm r}(\bm{x})-(\bm{\ddot{x}}_1+\bm{\dot{x}}_1)
\]
which serves as the new $\bm{x}_0$ for next function. Meanwhile, the server has formed its corresponding share, $(\bm{\ddot{x}}_1+\bm{\dot{x}}_1)$, at offline since $\bm{\ddot{x}}_1$ and $\bm{\dot{x}}_1$ are input-independent, and it serves as the new $\bm{x}_1$ for next function.

To reduce the added waiting cost caused by a hard queue jumping, PrivQJ explores a jigsaw-style computation within a batch such that any available slots in the intermediate ciphertext of in-queue inputs are recycled to produce unique pieces of output for prior inputs. Since the intermediate ciphertext mainly associate with \fbox{$\bm{t}$} at online and \doublebox{$\widehat{\bm{r}_0}$} at offline, as shown in Eq.~\eqref{eq:fi1}, we describe below how PrivQJ fully utilizes the peripheral slots and minimizes the batch size to enbale an efficient queue jumping for prior inputs.

\subsection{Full Slot Recycling in Cipher of \textit{\textbf{t}}}
\label{slot:recycle}
Recall from Eq.\eqref{eq:unfold:v3} that the cipher of \textit{\textbf{t}}, \fbox{$\bm{t}$}, is drived from \fbox{$\bm{h}_1$} and \fbox{$\bm{x}_1\times(\bm{1}-\bm{2}\times\bm{h}_1)$}, and the HE computation that associates with \fbox{$\bm{h}_1$} and \fbox{$\bm{x}_1\times(\bm{1}-\bm{2}\times\bm{h}_1)$} works completely in an element-wise manner. Thus the values in $\bm{h}_1$ and $\{\bm{x}_1\times(\bm{1}-\bm{2}\times\bm{h}_1)\}$ are tightly-packed and encrypted by $\mathcal{S}$ into \fbox{$\bm{h}_1$} and \fbox{$\bm{x}_1\times(\bm{1}-\bm{2}\times\bm{h}_1)$}, respectively. Here, one ciphertext slot is filled with one element from $\bm{h}_1$ or $\{\bm{x}_1\times(\bm{1}-\bm{2}\times\bm{h}_1)\}$, and there are thus $\lceil\frac{C_iH_iW_i}{N}\rceil$ ciphertext to encrypt $\bm{h}_1$ or $\{\bm{x}_1\times(\bm{1}-\bm{2}\times\bm{h}_1)\}$. 

Given that we always have $C_iH_iW_i\%N\neq0$, it leaves the tail slots in last ciphertext of $\bm{h}_1$ and $\{\bm{x}_1\times(\bm{1}-\bm{2}\times\bm{h}_1)\}$ idle. For example, there are 7168 idle slots in the last ciphertext under the general setting of $N=8192$ when encrypting one of the main blocks in ResNet, with 256 input channels each of which is in size $14\times14$~\cite{he2016deep}. This indicates a utilization rate around $12\%$ for that tail ciphertext. Those idle slots lead to a waste of computation and communication since all slots in a ciphertext are involved in each HE operation or ciphertext transmission.

While that low utilization rate seems unavoidable, we find that encrypting $\bm{h}_1$ and \{$\bm{x}_1\times(\bm{1}-\bm{2}\times\bm{h}_1)$\} for least multiple in-queue inputs enables \textit{an almost-free computation for prior inputs}. Such computation is realized by fully filling idle slots in all tail ciphertext of \fbox{$\bm{h}_1$} and \fbox{$\bm{x}_1\times(\bm{1}-\bm{2}\times\bm{h}_1)$} of in-queue inputs such that elements in all filled slots exactly form $\bm{h}_1$ and \{$\bm{x}_1\times(\bm{1}-\bm{2}\times\bm{h}_1)$\} for multiple prior inputs.

Specifically, let $\widehat{s}=(\lceil\frac{C_iH_iW_i}{N}\rceil{N}-C_iH_iW_i)$ be the number of idle slots in the tail ciphertext.
Then, the client is able to additionally obtain \fbox{$\bm{t}$} for $\widehat{s}/g(\widehat{s},C_iH_iW_i)$ prior inputs after normally computing the \fbox{$\bm{t}$} for $C_iH_iW_i/g(\widehat{s},C_iH_iW_i)$ in-queue inputs. 
Here $g(x,y)$ is the \textit{greatest common divisor} between $x$ and $y$, and the underlying logic is to fill the last ciphertext of \fbox{$\bm{h}_1$} and \fbox{$\{\bm{x}_1\times(\bm{1}-\bm{2}\times\bm{h}_1)\}$} for $C_iH_iW_i/g(\widehat{s},C_iH_iW_i)$ in-queue inputs with $\widehat{s}$ values from the $\bm{h}_1$ and $\{\bm{x}_1\times(\bm{1}-\bm{2}\times\bm{h}_1)\}$ for $\widehat{s}/g(\widehat{s},C_iH_iW_i)$ prior inputs. 
We call such correlated data assignment as chained batching where the encryption of $\bm{h}_1$ and $\{\bm{x}_1\times(\bm{1}-\bm{2}\times\bm{h}_1)\}$ for $C_iH_iW_i/g(\widehat{s},C_iH_iW_i)$ {in-queue inputs} subsequently enables free computation of \fbox{$\bm{t}$} for $\widehat{s}/g(\widehat{s},C_iH_iW_i)$ {prior inputs}, and the optimal batch size for a prior input is then determined as $C_iH_iW_i/\widehat{s}$.

\begin{figure}[!t]
\centering
\includegraphics[trim={6.5cm 8.8cm 10.9cm 1.5cm}, clip, scale=0.51]{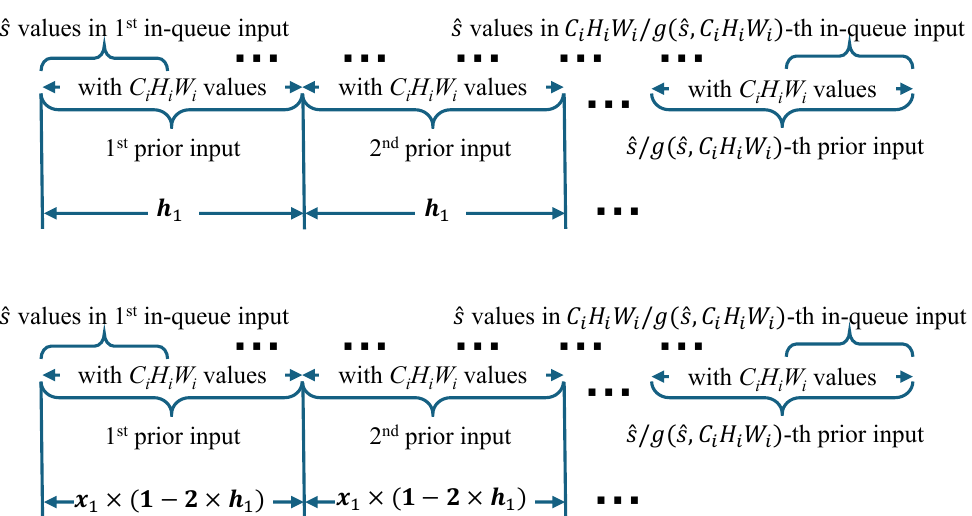}
\caption[Tail packing of h1 and x1]{Tail packing of \fbox{$\bm{h}_1$} and \fbox{$\{\bm{x}_1\times(\bm{1}-\bm{2}\times\bm{h}_1)\}$}.}
\label{fig:conv1}
\end{figure}
Figure~\ref{fig:conv1} shows the concrete data assignment for original in-queue inputs and emergent prior inputs. Specifically, given the offline-received and slot-recycled \fbox{$\bm{h}_1$} and \fbox{$\bm{x}_1\times(\bm{1}-\bm{2}\times\bm{h}_1)$} for each in-queue input at the client, the online process works as follows.
First, $\mathcal{C}$ feeds the freshly-obtained $\bm{x}_0$ and $\mathcal{S}$  feeds the pre-generated $\bm{x}_1$ and $\bm{h}_1$ to get involved in OT-based $f'_{\mathrm r}(\bm{x})$, e.g.,~\cite{rathee2020cryptflow2}. After that $\mathcal{C}$ obtains $\bm{h}_0$ namely the share for $f'_{\mathrm r}(\bm{x})$. Next, $\mathcal{C}$ computes \fbox{$\bm{t}$} according to Eq.~\eqref{eq:unfold:v3} by HE cMult and HE Add. 
Then, \fbox{$\bm{t}$} is sent to $\mathcal{S}$ which decrypts \fbox{$\bm{t}$} into $\bm{t}$, gets $(\bm{x}_1\times\bm{h}_1+\bm{t})$, and obtains the first convolution in Eq.~\eqref{eq:fi1} namely $\bm{y}=(\bm{x}_1\times\bm{h}_1+\bm{t})\ast\bm{k}$. 

Note that there are $\widehat{s}$ values, which are from the $\bm{h}_1$ for prior inputs, in the tail ciphertext of \fbox{$\bm{h}_1$} for each in-queue input, and it also works for the tail ciphertext of \fbox{$\bm{x}_1\times(\bm{1}-\bm{2}\times\bm{h}_1)$} for each in-queue input. 
As such, there are $\widehat{s}$ values, which are from $\bm{t}$ of prior inputs, in the $\mathcal{S}$-decrypted plaintext $\bm{t}\in\mathbb{Z}_p^{C_i\times H_i\times W_i+\widehat{s}}$ of each in-queue input, and the convolution $(\bm{x}_1\times\bm{h}_1+\bm{t})\ast\bm{k}$ is performed by $\mathcal{S}$ between kernel $\bm{k}$ and the first $C_iH_iW_i$ values from that plaintext $\bm{t}$.
At the same time, the last $\widehat{s}$ values from the same $\bm{t}$ are left aside to form a complete $\bm{t}$ for a prior input and the complete $\bm{t}$ is then convolved with kernel $\bm{k}$ to get the corresponding $\bm{y}$. Quantitatively, the complete $\bm{t}$ for one prior input is obtained after computing for $\lceil C_iH_iW_i/\widehat{s}\rceil$ in-queue inputs.



\begin{figure*}[!tbp]
\centering
\includegraphics[trim={0.2cm 3.8cm 0.1cm 2.2cm}, clip, scale=0.485]{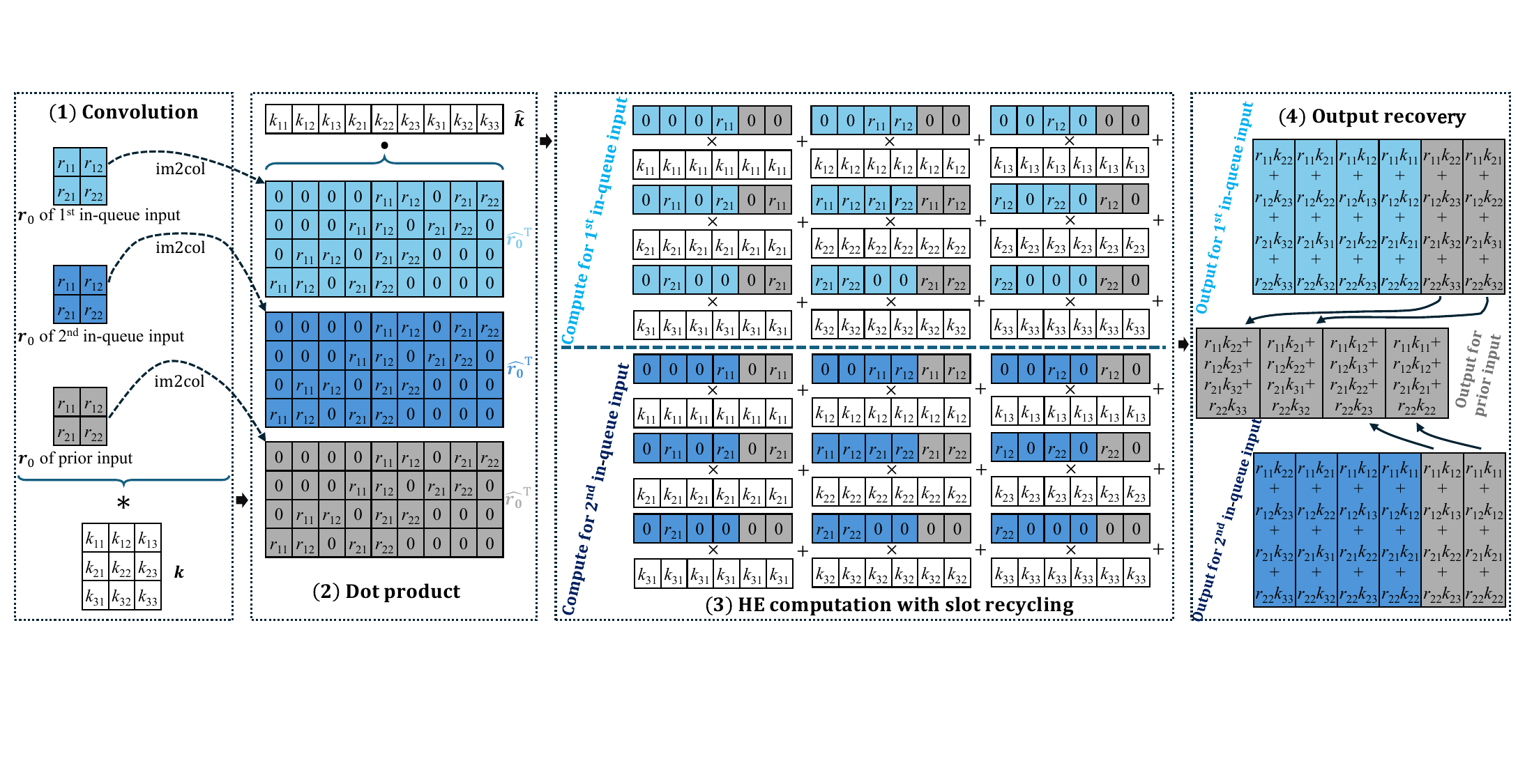}
\caption{Relationship among convolution, dot product, and HE computation with slot recycling.}
\label{fig:offline}
\end{figure*}

\begin{figure}[!tbp]
\centering
\includegraphics[trim={3.7cm 3.6cm 3.2cm 5cm}, clip, scale=0.54]{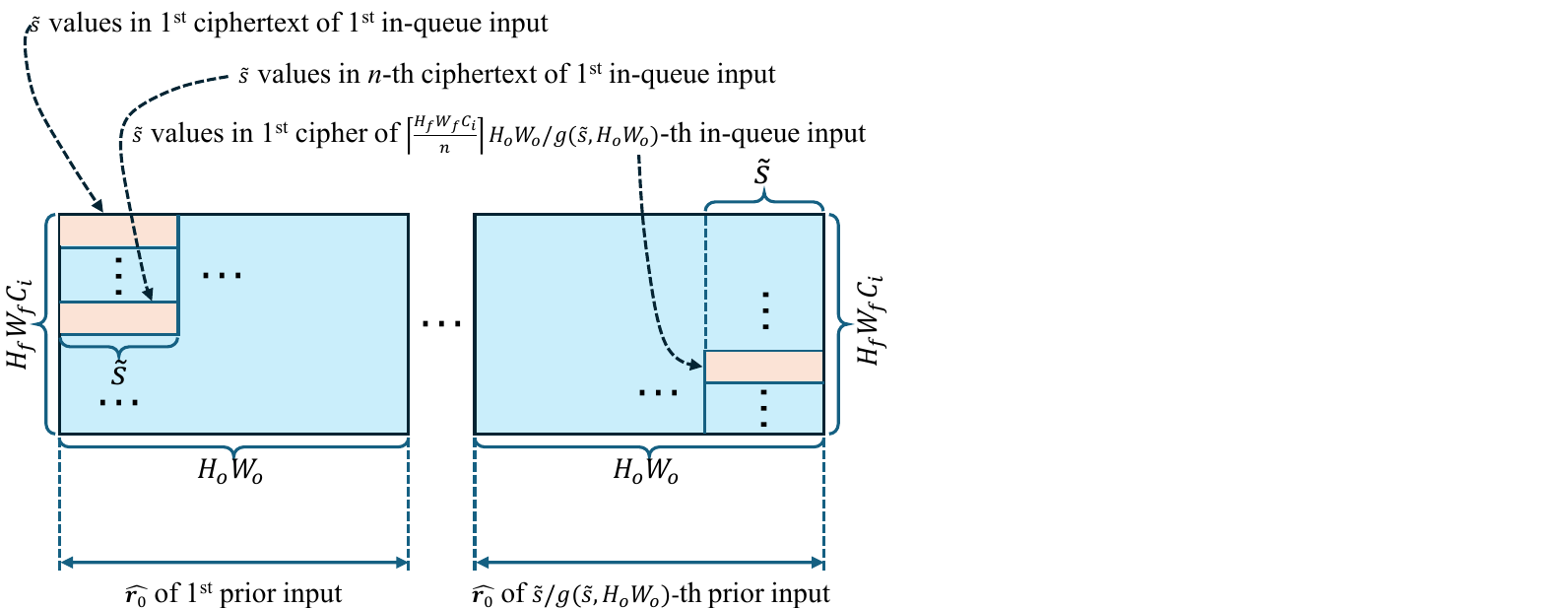}
\caption{Tail packing in cipher of $\widehat{\bm{r}_0}$ for in-queue inputs.}
\label{fig:conv2}
\end{figure}
\subsection{Full Slot Recycling in Cipher of \texorpdfstring{$\widehat{\textit{\textbf{r}}_0}$}{r0}}\label{r0kslotrec}

Recall that the convolution between $\bm{r}_0$ at $\mathcal{C}$ and $\bm{k}$ at $\mathcal{S}$ is first converted into the matrix-matrix multiplication between transformed $\bm{r}_0$, $\widehat{\bm{r}_0}\in\mathbb{Z}_p^{H_fW_fC_i\times H_oW_o}$, and flattened $\bm{k}$, $\widehat{\bm{k}}\in\mathbb{Z}_p^{C_o\times H_fW_fC_i}$, according to im2col~\cite{jia2014caffe}, and the $j$-th row $(j\in[C_o])$ in output of the matrix-matrix multiplication, $\widehat{\bm{k}}\cdot\widehat{\bm{r}_0}$, is viewed as the sum of all $H_fW_fC_i$ rows in $\widehat{\bm{r}_0}$, with $j_1$-th row $(j_1\in[H_fW_fC_i])$ scaled by the value at $j_1$-th column in $j$-th row of $\widehat{\bm{k}}$.
Then, that matrix-matrix multiplication is realized at offline by encrypting $\widehat{\bm{r}_0}$ at $\mathcal{C}$ and performing only HE cMult and Add at $\mathcal{S}$, as exemplified in Figure~\ref{fig:offline}.

Since \doublebox{$\widehat{\bm{r}_0}$} actually includes multiple ciphertext each of which packs multiple rows from $\widehat{\bm{r}_0}$ and each row of $\widehat{\bm{r}_0}$ is with length $H_oW_o$,
it leaves idle slots in the tail of each ciphertext when $N\%H_oW_o\gg1$, and such case always happens in neural models such as ResNet~\cite{he2016deep}.
Therefore, we propose to fully fill each ciphertext of \doublebox{$\widehat{\bm{r}_0}$} for the in-queue input with values that are chosen in row wise from $\widehat{\bm{r}_0}$ of prior inputs.
In this way, the convolution for \doublebox{$\widehat{\bm{r}_0}$} of each in-queue input remains its efficiency while part of the corresponding computation for prior inputs is performed simultaneously without any extra cost, and the complete convolution for each $\widehat{\bm{r}_0}$ of the prior input is obtained after normal computation for multiple in-queue inputs. In the following, we introduce the way to obtain the least number of in-queue inputs.

Quantitatively, recall that $\widehat{\bm{r}_0}$ is in size $H_fW_fC_i\times{H_oW_o}$ and the convolution, $\bm{r}_0\ast\bm{k}$, outputs $C_o$ channels such that the $j$-th channel $(j\in[C_o])$ is obtained by first weighting the $i$-th row $(i\in[H_fW_fC_i])$ of $\widehat{\bm{r}_0}$ with the $i$-th value in the $j$-th row of $\widehat{\bm{k}}$, and then summing the $H_fW_fC_i$ weighted rows. On the basis, $\lfloor\frac{N}{H_oW_o}\rfloor$ rows from $\widehat{\bm{r}_0}$ are packed in one ciphertext and $n=\lceil({H_fW_fC_i}/\lfloor\frac{N}{H_oW_o}\rfloor)\rceil$ ciphertext is formed at $\mathcal{C}$ to encrypt a complete $\widehat{\bm{r}_0}$.

Given $\Tilde{s}=N\%H_oW_o$ idle slots in each of $n$ ciphertext, $n$ rows from $\widehat{\bm{r}_0}$ of prior inputs, with each row in length $\Tilde{s}$, are respectively packed at the tail of $n$ ciphertext for in-queue inputs, and we thus need $\lceil(H_fW_fC_i/n)\rceil$ in-queue inputs to pack $\Tilde{s}$ columns from $\widehat{\bm{r}_0}$ of one prior input. As such encrypting $\widehat{\bm{r}_0}$ for $\lceil(H_fW_fC_i/n)\rceil{H_oW_o/g(\Tilde{s}, H_oW_o)}$ in-queue inputs enables the exact encryption of $\widehat{\bm{r}_0}$ for $\Tilde{s}/g(\Tilde{s}, H_oW_o)$ prior inputs, and it makes the computation of $\bm{r}_0\ast\bm{k}$ for those prior inputs in an almost-free manner. In other words, each ciphertext of \doublebox{$\widehat{\bm{r}_0}$} for in-queue inputs simultaneously packs values from $\widehat{\bm{r}_0}$ of prior inputs, and Figure~\ref{fig:conv2} shows the data assignment in detail. 

We recall Figure~\ref{fig:offline} to give a more concrete description for the full slot recycling in cipher of $\widehat{\textit{\textbf{r}}_0}$. Specifically, it demonstrates an example to compute $\bm{r}_0\ast\bm{k}$ for three inputs including a batch of two in-queue inputs and one emergent prior input. The three convolution is first converted into three dot product. Here each $\widehat{\bm{r}_0}$ is in transposed form and is obtained according to im2col, and we assume that each ciphertext contains six slots. 
Then, one row from $\widehat{\bm{r}_0}$ of in-queue input along with two values from one row of $\widehat{\bm{r}_0}$ for the prior input are packed and encrypted in one ciphertext to form \doublebox{$\widehat{\bm{r}_0}$} for in-queue inputs. 

Here the encryption for $\widehat{\bm{r}_0}$ is performed by $\mathcal{C}$, and the corresponding encryption for prior input is completely shared within the encryption for in-queue inputs without any other overhead. Upon receiving \doublebox{$\widehat{\bm{r}_0}$}, $\mathcal{S}$ performs HE cMult between the plaintext kernel and \doublebox{$\widehat{\bm{r}_0}$}, and sums up the multiplied ciphertext by HE Add. In this way, one added ciphertext corresponds to one of $C_o$ output channels, and $C_o$ added ciphertext is obtained for each in-queue input and is masked with randomness to send back to $\mathcal{C}$. Then, $\mathcal{C}$ decrypts masked ciphertext to get it's share of $\bm{r}_0\ast\bm{k}$ for all three inputs, while the randomness for masking forms the share at $\mathcal{S}$.
Note that such process is completed at \textit{offline} and $\bm{r}_0\ast\bm{k}$ for each prior input is obtained after computing for $\lceil(H_fW_fC_i/n)\rceil{H_oW_o}/\Tilde{s}$ in-queue inputs.

\subsection{Putting All Together to Build PrivQJ}
Let's now systematize the overall computation of PrivQJ which enables a much more efficient queue jumping. As illustrated in Figure~\ref{fig:reluconv}, there are two phases namely offline and online. The offline involves two one-way transmissions respectively from the server and the client. More concretely, $\mathcal{S}$ encrypts $\bm{h}_1$ and $\{\bm{x}_1\times(\bm{1}-\bm{2}\times\bm{h}_1)\}$ with slot recycling shown in Figure~\ref{fig:conv1}, while $\mathcal{C}$ encrypts $\widehat{\bm{r}_0}$ with slot recycling shown in Figure~\ref{fig:conv2} and then interacts with $\mathcal{S}$ to share $\bm{r}_0\ast\bm{k}$.
Here, a pair of values, i.e., the number of in-queue inputs in a batch and the one of prior inputs, are determined during the encryption at $\mathcal{S}$ and $\mathcal{C}$, respectively. Meanwhile, the resultant two pairs of values do not need to be the same since they could work independently at offline to form a pool of ciphertext and shares for the online queue jumping.

At online, $\mathcal{S}$ and $\mathcal{C}$ first collaborate to compute $\bm{h}_0$, i.e., $\mathcal{C}$'s share of $f'_{\mathrm{r}}(\bm{x})$, via OT-based protocol using pregenerated $\bm{x}_1$ and $\bm{h}_1$ at $\mathcal{S}$ and freshly-obtained $\bm{x}_0$ at $\mathcal{C}$. 
Here $\bm{x}_1$, $\bm{h}_1$, $\bm{x}_0$, and $\bm{h}_0$ are formed in a slot-recycled manner, with $C_i{H_i}{W_i}$ elements for one in-queue input and $\widehat{s}$ values for the prior input. 
After $\mathcal{C}$ obtains $\bm{h}_0$, \fbox{$\bm{t}$} is computed according to Eq.~\eqref{eq:unfold:v3} and is then sent to $\mathcal{S}$. $\mathcal{S}$ decrypts \fbox{$\bm{t}$} into $\bm{t}$, and divides the whole $\bm{t}$ into two parts: One with $C_i{H_i}{W_i}$ elements for an in-queue input, and the other with $\widehat{s}$ values for prior input.
Next, $\mathcal{S}$ computes $(\bm{x}_1\times\bm{h}_1+\bm{t})\ast\bm{k}$ namely $\bm{y}$ based on the $\bm{t}$ for in-queue input, and $\bm{y}$ is masked with pre-generated randomness $\bm{\dot{x}}_1$ as $(\bm{y}-\bm{\dot{x}}_1)$ and is then sent back to the client. Till now, $\mathcal{C}$ forms a new $\bm{x}_0$ as its share for next function based on $(\bm{y}-\bm{\dot{x}}_1)$ and offline-received share for $\bm{r}_0\ast\bm{k}$, while $\mathcal{S}$ forms its new $\bm{x}_1$ at offline based on pre-generated masks for 
$\bm{r}_0\ast\bm{k}$ and $\bm{y}$.

At the same time, once $\mathcal{S}$ collects all values of $\bm{t}$ for prior inputs, corresponding $\bm{y}$ namely $(\bm{x}_1\times\bm{h}_1+\bm{t})\ast\bm{k}$ is computed. Then $\mathcal{S}$ masks $\bm{y}$ with randomness for prior inputs, $\bm{\dot{x}}_1$, and $(\bm{y}-\bm{\dot{x}}_1)$ is sent to $\mathcal{C}$ to similarly form the input shares of next function. Note that the online involves $\widehat{s}/g(\widehat{s},C_iH_iW_i)$ {prior inputs} and $C_iH_iW_i/g(\widehat{s},C_iH_iW_i)$ in-queue inputs. Since a neural model has multiple blocks of $f_{\mathrm c}f_{\mathrm r}(\bm{x})$, one strategy to select the number of in-queue inputs, i.e., the batch size, for the model is the largest batch size for a prior input among all blocks namely the largest $C_iH_iW_i/\widehat{s}$, and that number could be further optimized by decreasing $C_i$, $H_i$, and $W_i$ through input segmentation.

Furthermore, a neural model contains other functions than $f_{\mathrm c}f_{\mathrm r}(\bm{x})$ and some model adjustments are applied to make full use of proposed module, and more details are explained in Appendix.

\begin{table*}[!tb]\color{black}
\centering
\scriptsize
\caption{Quantitative comparison of waiting cost added to in-queue inputs that follow rightly after the prior inputs.
}\label{complexity:fcfr}
\resizebox{\textwidth}{!}{%
\begin{tabular}{c | c c c c | c c c c c c c  }
\hline\hline
\multirow{2}{*}{Frameworks}& \multicolumn{4}{c|}{Offline Computation} & \multicolumn{7}{c}{Online Computation} \\
\cline{2-12}
& \# Enc &\# cMult & \# Dec & \# Add & Mx & \# Rot & \# Enc & \# cMult & \# Dec & \# Add & \# Extr\\
\hline\hline
CrypTFlow2~\cite{rathee2020cryptflow2}& - & - & - & - & \checkmark & $\geq\frac{(f_h^2-1)C_i}{C_n}+C_o-\frac{C_o}{C_n}$ & $\frac{C_i}{C_n}$ & $\frac{f_h^2C_iC_o}{C_n}$ & $\frac{C_o}{C_n}$ & $\frac{C_i+C_oC_if_h^2}{C_n}$ & -\\
\hline
Cheetah~\cite{huang2022cheetah}& - & - & - & - & \checkmark & - & $ \frac{C_iH_iW_i}{N} $ & $ \frac{C_oC_iH_iW_i}{N} $ & $ C_oH_oW_o $ & $C_oH_oW_o $ & $ C_oH_oW_o$\\
\hline
{FIT}~\cite{qiaosp24}& $\frac{f_h^2C_i}{C_n}$ &$\frac{f_h^2C_iC_o}{C_n}$ & $C_o$ & $\frac{f_h^2C_iC_o}{C_n}$ & \crossmark & - & - & $2\frac{C_i}{C_n}$ & $\frac{C_i}{C_n}$ & $2\frac{C_i}{C_n}$& -\\
\hline
{NEXUS}~\cite{cryptoeprint:2024/136}$^*$& $C_o$ &${f_h^2C_iC_o}$ & - & ${(f_h^2C_i-1)C_o}$ & \checkmark & - & - & - & $C_o$ & $2C_o$& -\\
\hline
\textbf{PrivQJ} (this work)& - &- & - & - & \crossmark & - & - & - & - & -& -\\
\hline\hline
\multicolumn{12}{l}{$C_n=\lfloor \frac{N}{C_iH_iW_i}\rfloor$. $^*$ Tailored for mixed-primitive inference. }
\end{tabular}
}
\end{table*}

\subsection{Complexity and Security}
\label{sys:complexity}
In this section, we analyze the complexity and security of proposed \textrm{PrivQJ}, aiming to provide quantitative results to verify the efficiency of proposed queue jumping as well as the privacy preservation throughout the whole process. In general, the efficiency includes the complexity for both in-queue input and prior one, and the privacy preservation is based on concrete security proof.

\textbf{Complexity}. The complexity of \textrm{PrivQJ} focuses on the added waiting cost to the in-queue inputs which are forced to follow after the prior one when the queue jumping happens, and Table~\ref{complexity:fcfr} demonstrates concrete numbers of each crypto operations. Here, the computation is for $f_{\mathrm c}f_{\mathrm r}(\bm{x})$ and the unique complexity advantages of \textrm{PrivQJ} lie in two folds. 
First, the computation for each in-queue input is tailored from FIT framework with less computation cost, as analyzed in Section~\ref{sys:mechanism}. Since FIT remains competitive in both offline and online computation as illustrated in Table~\ref{complexity:fcfr}, the efficiency for in-queue inputs is guaranteed.
Second, the efficiency for computing each prior input is almost-free without introducing any extra HE cost. Such ultralight computation is realized by our in-processing slot recycling that fully embeds computation for prior inputs into the computing process for in-queue inputs.

\textbf{Security.} We rely on ideal/real-world paradigm~\cite{howtosimulate} to prove the security of \textrm{PrivQJ}, and the details are explained in Appendix.

\section{Evaluation}\label{eval}
We implement \textrm{PrivQJ} based on code from CrypTFlow2~\cite{rathee2020cryptflow2} and the traffic control command in Linux is utilized to simulate different networking conditions, and our code is available \href{https://anonymous.4open.science/r/PrivQJ-13C2}{\textit{here}}. Specifically, the Local Area Network (LAN) is with bandwidth about 3 gigabit per second (3Gbps) and ping time about 0.8ms, and the Wide Area Networks (WANs) are configured in four settings with WAN$_1$ in 100 megabit per second (100Mbps) bandwidth and 40ms ping time, WAN$_2$ in 100Mbps and 80ms ping time, WAN$_3$ in 200Mbps and 40ms ping time, and WAN$_4$ in 200Mbps and 80ms ping time. These settings are inline with state-of-the-art systems \cite{pangsp24,cryptoeprint:2024/136} and
the machine is with Intel(R) Core$^{\textrm{TM}}$ i9-14900KF$\times$32 CPU and 128GB of RAM, and all tests run with one thread.

In the following, we first evaluate performance of queue jumping when computing basic blocks in popular CNNs for ImageNet dataset \cite{imagenet}, with concrete discussion on the offline and online efficiency among different approaches, i.e., CrypTFlow2~\cite{rathee2020cryptflow2}, Cheetah \cite{huang2022cheetah}, and FIT \cite{qiaosp24}. Then, the efficiency of PrivQJ to compute in-queue inputs is compared with others to demonstrate the competitive performance for normal computation. Finally, we apply PrivQJ into well-known neural models such as ResNet \cite{he2016deep}, VGG \cite{simonyan2014very} and AlexNet \cite{krizhevsky2012imagenet} to show its advantages over state-of-the-art works.

\begin{table*}
    \centering
    \footnotesize
    \caption{Overall cost added to in-queue input after prior one.
}\label{eval::basicblock}
    \begin{tabular}{c|cccccc}
    \hline
    \multicolumn{7}{c}{Communication cost (comm.) \textit{in MiB} and time cost \textit{in second}}\\
    \hline
    \multicolumn{3}{c}{$H_i,C_i,f_h,C_o$}& 56,64,3,64 & 28,128,3,128 & 14,256,3,256 & 7,512,3,512\\
    \hline
    \multirow{16}{*}{\rotatebox{90}{\makebox[0pt][c]{LAN}}}& \multirow{4}{*}{\cite{rathee2020cryptflow2}}& comm.& 228.1 & 115.8 & 59.1 & 32.1 \\
     & & speedup& \textbf{152}$\times$& \textbf{152}$\times$& \textbf{155}$\times$& \textbf{168}$\times$\\
    \cline{3-3}
     & & time& 22.1& 20.6& 20.2&39.6 \\
     & & speedup& \textbf{315}$\times$& \textbf{343}$\times$& \textbf{404}$\times$& \textbf{792}$\times$\\
    \cline{2-7}
    &\multirow{4}{*}{\cite{huang2022cheetah}$^*$}& comm.& 14.1& 9.9& 10.8& 17.2\\
     & & speedup& \textbf{9}$\times$& \textbf{13}$\times$& \textbf{28}$\times$&\textbf{90}$\times$ \\
    \cline{3-3}
     & & time& 1.4& 1.2& 1.2& 1.77\\
     & & speedup& \textbf{20}$\times$& \textbf{20}$\times$&\textbf{24}$\times$ &\textbf{35}$\times$ \\
    \cline{2-7}
    &\multirow{4}{*}{\cite{qiaosp24}}& comm.& 295.9& 154.7& 102.4& 99.9\\
     & & speedup& \textbf{197}$\times$& \textbf{203}$\times$& \textbf{269}$\times$&\textbf{525}$\times$ \\
    \cline{3-3}
     & & time& 15.5&11.8 & 11.1& 11\\
     & & speedup& \textbf{221}$\times$& \textbf{196}$\times$& \textbf{222}$\times$&\textbf{220}$\times$ \\
    \cline{2-7}
    &\multirow{4}{*}{\rotatebox{90}{\makebox[0pt][c]{\textbf{PrivQJ}}}}& \multirow{2}{*}{comm.}& \multirow{2}{*}{1.5}& \multirow{2}{*}{0.76}&  \multirow{2}{*}{0.38}& \multirow{2}{*}{0.19}\\
     & & & & & & \\
     \cline{3-3}
     & & \multirow{2}{*}{time}& \multirow{2}{*}{0.07}& \multirow{2}{*}{0.06}& \multirow{2}{*}{0.05}& \multirow{2}{*}{0.05}\\
      & & & & & & \\     
    \hline
    \multirow{9}{*}{\rotatebox{90}{\makebox[0pt][c]{WAN$_1$}}}& \multirow{2}{*}{\cite{rathee2020cryptflow2}}& time& 41.7& 30.7& 25.5& 42.9\\
     & & speedup& \textbf{595}$\times$& \textbf{495}$\times$& \textbf{500}$\times$&\textbf{875}$\times$ \\
    \cline{2-7}
    &\multirow{2}{*}{\cite{huang2022cheetah}$^*$}& time& 2.9& 2.4& 2.5& 3.7\\
     & & speedup& \textbf{41}$\times$& \textbf{38}$\times$& \textbf{49}$\times$&\textbf{75}$\times$ \\
    \cline{2-7}
    &\multirow{2}{*}{\cite{qiaosp24}}& time& 40.7& 25.3&20.5 & 20.1\\
     & & speedup& \textbf{581}$\times$& \textbf{408}$\times$& \textbf{401}$\times$&\textbf{410}$\times$ \\
    \cline{2-7}
    &\multirow{3}{*}{\rotatebox{90}{\makebox[0pt][c]{\textbf{PrivQJ}}}}& \multirow{3}{*}{time}& \multirow{3}{*}{0.07}& \multirow{3}{*}{0.062}& \multirow{3}{*}{0.051}& \multirow{3}{*}{0.049}\\
     & & & & & & \\
      & & & & & & \\     
    \hline
    \multirow{9}{*}{\rotatebox{90}{\makebox[0pt][c]{WAN$_2$}}}& \multirow{2}{*}{\cite{rathee2020cryptflow2}}& time& 44.7& 32.9& 27.3& 44.1\\
     & & speedup& \textbf{629}$\times$& \textbf{514}$\times$& \textbf{455}$\times$& \textbf{900}$\times$\\
    \cline{2-7}
    &\multirow{2}{*}{\cite{huang2022cheetah}$^*$}& time& 3.9& 2.9& 3&3.9 \\
     & & speedup& \textbf{54}$\times$& \textbf{45}$\times$& \textbf{57}$\times$& \textbf{79}$\times$\\
    \cline{2-7}
    &\multirow{2}{*}{\cite{qiaosp24}}& time& 44.6& 28.1& 22.3&21.8 \\
     & & speedup& \textbf{628}$\times$& \textbf{439}$\times$& \textbf{428}$\times$& \textbf{444}$\times$\\
    \cline{2-7}
    &\multirow{3}{*}{\rotatebox{90}{\makebox[0pt][c]{\textbf{PrivQJ}}}}& \multirow{3}{*}{time}&\multirow{3}{*}{0.071} &\multirow{3}{*}{0.064} &\multirow{3}{*}{0.052} &\multirow{3}{*}{0.049} \\
     & & & & & & \\
      & & & & & & \\     
    \hline
    \multirow{9}{*}{\rotatebox{90}{\makebox[0pt][c]{WAN$_3$}}}& \multirow{2}{*}{\cite{rathee2020cryptflow2}}& time& 32.8& 26.6& 23.5& 41.8\\
     & & speedup& \textbf{449}$\times$& \textbf{350}$\times$& \textbf{385}$\times$& \textbf{853}$\times$\\
    \cline{2-7}
    &\multirow{2}{*}{\cite{huang2022cheetah}$^*$}& time& 3& 2.3&2.1 &2.9 \\
     & & speedup&\textbf{41}$\times$ & \textbf{30}$\times$& \textbf{34}$\times$& \textbf{59}$\times$\\
    \cline{2-7}
    &\multirow{2}{*}{\cite{qiaosp24}}& time& 29.2&19.7 &16.4 &16.3 \\
     & & speedup& \textbf{400}$\times$& \textbf{259}$\times$& \textbf{268}$\times$& \textbf{332}$\times$\\
    \cline{2-7}
    &\multirow{3}{*}{\rotatebox{90}{\makebox[0pt][c]{\textbf{PrivQJ}}}}& \multirow{3}{*}{time}& \multirow{3}{*}{0.073}& \multirow{3}{*}{0.076}& \multirow{3}{*}{0.061}&\multirow{3}{*}{0.049} \\
     & & & & & & \\
      & & & & & & \\     
    \hline
    \multirow{9}{*}{\rotatebox{90}{\makebox[0pt][c]{WAN$_4$}}}& \multirow{2}{*}{\cite{rathee2020cryptflow2}}& time& 35.8&28.9 &25 &43.1 \\
     & & speedup& \textbf{464}$\times$& \textbf{370}$\times$& \textbf{446}$\times$& \textbf{862}$\times$\\
    \cline{2-7}
    &\multirow{2}{*}{\cite{huang2022cheetah}$^*$}& time& 2.8&2.6 &2.6 &3.4 \\
     & & speedup& \textbf{36}$\times$&\textbf{33}$\times$ &\textbf{46}$\times$ &\textbf{68}$\times$ \\
    \cline{2-7}
    &\multirow{2}{*}{\cite{qiaosp24}}& time& 32.9& 21.8& 18.4& 18\\
     & & speedup& \textbf{427}$\times$& \textbf{279}$\times$& \textbf{328}$\times$&\textbf{360}$\times$ \\
    \cline{2-7}
    &\multirow{3}{*}{\rotatebox{90}{\makebox[0pt][c]{\textbf{PrivQJ}}}}& \multirow{3}{*}{time}& \multirow{3}{*}{0.077}& \multirow{3}{*}{0.078}& \multirow{3}{*}{0.056}&\multirow{3}{*}{0.05} \\
     & & & & & & \\
      & & & & & & \\     
    \hline
    \multicolumn{7}{l}{$^*$ with more efficient HE and OT backends.}\\
    \end{tabular}
\end{table*}
\subsection{Added Cost When Computing Basic Blocks}
Table \ref{eval::basicblock} demonstrates the overall waiting cost, including the offline and online overhead, added to in-queue input that follows after the prior one when computing basic blocks in the neural networks. Specifically, PrivQJ shows its unique advantage that it just adds little cost to the in-queue input, with cost reduction from one to two orders of magnitudes, in both communication and computation time. The fundamental reason lies in the proposed in-processing computation that embeds most cost for prior input into the computation for in-queue inputs within in a batch, leading to an almost-free process for the prior input. In contrast, other works need to run another new computation for prior input, which results in the non-trivial cost added to the subsequent in-queue inputs. Note that the added time cost when applying PrivQJ is always small in various networking settings as the communication volume is light and we further merge the transmission of the final sharing for prior input into the one for the last in-queue input in the batch namely we merge the final sharing for in-queue inputs and the final one for prior input in Figure \ref{fig:reluconv}, effectively mitigating the variation of added time in different round-trip settings.

Since the offline is input-independent and the online cost is more critical, we then evaluate the added cost at running time namely the online stage, and Table \ref{eval::online} shows the corresponding comparison. Key takeaway is that PrivQJ keeps its advantage in adding little cost to in-queue input even though the FIT framework \cite{qiaosp24} features with much more efficient running-time performance especially in LAN, where PrivQJ still reduces the added cost by 4 to 21 times compared with FIT. Furthermore, while cheetah \cite{huang2022cheetah} is with more efficient HE and OT backends and always has lower cost compared with CrypTFlow2 \cite{rathee2020cryptflow2} and FIT, the proposed PrivQJ is capable of further reducing the added cost by an order of magnitude, which mainly attributes to the effective elimination of HE cost and other communication overhead for prior input.

\begin{table*}
    \centering
    \footnotesize
    \caption{Online cost added to in-queue input after prior one.
}\label{eval::online}
    \begin{tabular}{c|cccccc}
    \hline
    \multicolumn{7}{c}{Communication cost (comm.) \textit{in MiB} and time cost \textit{in second}}\\
    \hline
    \multicolumn{3}{c}{$H_i,C_i,f_h,C_o$}& 56,64,3,64 & 28,128,3,128 & 14,256,3,256 & 7,512,3,512\\
    \hline
    \multirow{16}{*}{\rotatebox{90}{\makebox[0pt][c]{LAN}}}& \multirow{4}{*}{\cite{rathee2020cryptflow2}}& comm.& 228.1 & 115.8 & 59.1 & 32.1 \\
     & & speedup& \textbf{152}$\times$& \textbf{152}$\times$& \textbf{155}$\times$& \textbf{168}$\times$\\
    \cline{3-3}
     & & time& 22.1& 20.6& 20.2&39.6 \\
     & & speedup& \textbf{315}$\times$& \textbf{343}$\times$& \textbf{404}$\times$& \textbf{792}$\times$\\
    \cline{2-7}
    &\multirow{4}{*}{\cite{huang2022cheetah}$^*$}& comm.& 14.1& 9.9& 10.8& 17.2\\
     & & speedup& \textbf{9}$\times$& \textbf{13}$\times$& \textbf{28}$\times$&\textbf{90}$\times$ \\
    \cline{3-3}
     & & time& 1.4& 1.2& 1.2& 1.77\\
     & & speedup& \textbf{20}$\times$& \textbf{20}$\times$&\textbf{24}$\times$ &\textbf{35}$\times$ \\
    \cline{2-7}
    &\multirow{4}{*}{\cite{qiaosp24}}& comm.& 203.4& 103.2& 52.6& 26.9\\
     & & speedup& \textbf{135}$\times$& \textbf{135}$\times$& \textbf{138}$\times$&\textbf{141}$\times$ \\
    \cline{3-3}
     & & time& 1.5&0.8 & 0.4& 0.2\\
     & & speedup& \textbf{21}$\times$& \textbf{13}$\times$& \textbf{8}$\times$&\textbf{4}$\times$ \\
    \cline{2-7}
    &\multirow{4}{*}{\rotatebox{90}{\makebox[0pt][c]{\textbf{PrivQJ}}}}& \multirow{2}{*}{comm.}& \multirow{2}{*}{1.5}& \multirow{2}{*}{0.76}&  \multirow{2}{*}{0.38}& \multirow{2}{*}{0.19}\\
     & & & & & & \\
     \cline{3-3}
     & & \multirow{2}{*}{time}& \multirow{2}{*}{0.07}& \multirow{2}{*}{0.06}& \multirow{2}{*}{0.05}& \multirow{2}{*}{0.05}\\
      & & & & & & \\     
    \hline
    \multirow{9}{*}{\rotatebox{90}{\makebox[0pt][c]{WAN$_1$}}}& \multirow{2}{*}{\cite{rathee2020cryptflow2}}& time& 41.7& 30.7& 25.5& 42.9\\
     & & speedup& \textbf{595}$\times$& \textbf{495}$\times$& \textbf{500}$\times$&\textbf{875}$\times$ \\
    \cline{2-7}
    &\multirow{2}{*}{\cite{huang2022cheetah}$^*$}& time& 2.9& 2.4& 2.5& 3.7\\
     & & speedup& \textbf{41}$\times$& \textbf{38}$\times$& \textbf{49}$\times$&\textbf{75}$\times$ \\
    \cline{2-7}
    &\multirow{2}{*}{\cite{qiaosp24}}& time& 19.1& 10.1&5.5 & 3.1\\
     & & speedup& \textbf{272}$\times$& \textbf{162}$\times$& \textbf{107}$\times$&\textbf{63}$\times$ \\
    \cline{2-7}
    &\multirow{3}{*}{\rotatebox{90}{\makebox[0pt][c]{\textbf{PrivQJ}}}}& \multirow{3}{*}{time}& \multirow{3}{*}{0.07}& \multirow{3}{*}{0.062}& \multirow{3}{*}{0.051}& \multirow{3}{*}{0.049}\\
     & & & & & & \\
      & & & & & & \\     
    \hline
    \multirow{9}{*}{\rotatebox{90}{\makebox[0pt][c]{WAN$_2$}}}& \multirow{2}{*}{\cite{rathee2020cryptflow2}}& time& 44.7& 32.9& 27.3& 44.1\\
     & & speedup& \textbf{629}$\times$& \textbf{514}$\times$& \textbf{455}$\times$& \textbf{900}$\times$\\
    \cline{2-7}
    &\multirow{2}{*}{\cite{huang2022cheetah}$^*$}& time& 3.9& 2.9& 3&3.9 \\
     & & speedup& \textbf{54}$\times$& \textbf{45}$\times$& \textbf{57}$\times$& \textbf{79}$\times$\\
    \cline{2-7}
    &\multirow{2}{*}{\cite{qiaosp24}}& time& 21.9& 12.1& 6.8&4.1 \\
     & & speedup& \textbf{308}$\times$& \textbf{189}$\times$& \textbf{130}$\times$& \textbf{83}$\times$\\
    \cline{2-7}
    &\multirow{3}{*}{\rotatebox{90}{\makebox[0pt][c]{\textbf{PrivQJ}}}}& \multirow{3}{*}{time}&\multirow{3}{*}{0.071} &\multirow{3}{*}{0.064} &\multirow{3}{*}{0.052} &\multirow{3}{*}{0.049} \\
     & & & & & & \\
      & & & & & & \\     
    \hline
    \multirow{9}{*}{\rotatebox{90}{\makebox[0pt][c]{WAN$_3$}}}& \multirow{2}{*}{\cite{rathee2020cryptflow2}}& time& 32.8& 26.6& 23.5& 41.8\\
     & & speedup& \textbf{449}$\times$& \textbf{350}$\times$& \textbf{385}$\times$& \textbf{853}$\times$\\
    \cline{2-7}
    &\multirow{2}{*}{\cite{huang2022cheetah}$^*$}& time& 3& 2.3&2.1 &2.9 \\
     & & speedup&\textbf{41}$\times$ & \textbf{30}$\times$& \textbf{34}$\times$& \textbf{59}$\times$\\
    \cline{2-7}
    &\multirow{2}{*}{\cite{qiaosp24}}& time& 11&6.2 &3.5 &2.2 \\
     & & speedup& \textbf{150}$\times$& \textbf{81}$\times$& \textbf{57}$\times$& \textbf{44}$\times$\\
    \cline{2-7}
    &\multirow{3}{*}{\rotatebox{90}{\makebox[0pt][c]{\textbf{PrivQJ}}}}& \multirow{3}{*}{time}& \multirow{3}{*}{0.073}& \multirow{3}{*}{0.076}& \multirow{3}{*}{0.061}&\multirow{3}{*}{0.049} \\
     & & & & & & \\
      & & & & & & \\     
    \hline
    \multirow{9}{*}{\rotatebox{90}{\makebox[0pt][c]{WAN$_4$}}}& \multirow{2}{*}{\cite{rathee2020cryptflow2}}& time& 35.8&28.9 &25 &43.1 \\
     & & speedup& \textbf{464}$\times$& \textbf{370}$\times$& \textbf{446}$\times$& \textbf{862}$\times$\\
    \cline{2-7}
    &\multirow{2}{*}{\cite{huang2022cheetah}$^*$}& time& 2.8&2.6 &2.6 &3.4 \\
     & & speedup& \textbf{36}$\times$&\textbf{33}$\times$ &\textbf{46}$\times$ &\textbf{68}$\times$ \\
    \cline{2-7}
    &\multirow{2}{*}{\cite{qiaosp24}}& time& 13.9& 8& 4.8& 3.2\\
     & & speedup& \textbf{180}$\times$& \textbf{102}$\times$& \textbf{85}$\times$&\textbf{64}$\times$ \\
    \cline{2-7}
    &\multirow{3}{*}{\rotatebox{90}{\makebox[0pt][c]{\textbf{PrivQJ}}}}& \multirow{3}{*}{time}& \multirow{3}{*}{0.077}& \multirow{3}{*}{0.078}& \multirow{3}{*}{0.056}&\multirow{3}{*}{0.05} \\
     & & & & & & \\
      & & & & & & \\     
    \hline
    \multicolumn{7}{l}{$^*$ with more efficient HE and OT backends.}\\
    \end{tabular}
\end{table*}

\subsection{Performance for Each In-queue Input}

While PrivQJ features with much smaller cost added to subsequent in-queue inputs when the queue jumping happens, the efficiency for each in-queue input is also important since degrading the efficiency for in-queue inputs is equivalent with non-trivial waiting cost added to subsequent in-queue inputs. Therefore, Table \ref{eval::inqueue} compares the performance, including the offline plus online namely the overall cost and the running-time namely the online cost, of the in-queue input among different systems. Specifically, as for the overall cost in Table \ref{eval::inqueue}(a), PrivQJ has similar or slightly better time efficiency compared with FIT \cite{qiaosp24} which is more efficient than CrypTFlow2 \cite{rathee2020cryptflow2}. As for the communication cost, PrivQJ shows comparable overhead to FIT \cite{qiaosp24}. 

While the overall communication cost of each in-queue input in PrivQJ is relatively larger, a further look at the online performance in Table \ref{eval::inqueue}(b) indicates less cost than CrypTFlow2 \cite{rathee2020cryptflow2} and comparable cost to FIT \cite{qiaosp24}, and PrivQJ takes less online time in LAN compared to other systems. Moreover, cheetah \cite{huang2022cheetah} needs less online communication volume mainly due to it optimizes the OT module over CrypTFlow2, and the online communication and computation cost of PrivQJ is able to be further reduced to the similar level of cheetah by replacing the OT module for comparison with the one from cheetah, and by using the faster HE backend of cheetah. The quantitative reasons lie in two folds. First, PrivQJ only needs the OT-based computation for derivative of ReLU, without the following OT-based multiplexing that is required in cheetah. Second, PrivQJ is able to utilize \textit{every} slot in the ciphertext to encrypt the input and only share plaintext to form final shares while the sharing in cheetah is in ciphertext with more communication cost. Therefore, both the running-time computation and communication overhead of PrivQJ is similar with cheetah when shifting backends from CrypTFlow2 to cheetah.

\begin{table*}
    \centering
    \footnotesize
    \caption{Efficiency for individual in-queue input.
}\label{eval::inqueue}
    \begin{tabular}{c|cccccc}
    \hline
    \multicolumn{7}{c}{\textbf{(a)} \textit{Overall cost}, communication cost (comm.) \textit{in MiB} and time cost \textit{in second}}\\
    \hline
    \multicolumn{3}{c}{$H_i,C_i,f_h,C_o$}& 56,64,3,64 & 28,128,3,128 & 14,256,3,256 & 7,512,3,512\\
    \hline
    \multirow{10}{*}{\rotatebox{90}{\makebox[0pt][c]{LAN}}}& \multirow{2}{*}{\cite{rathee2020cryptflow2}}& comm.& 228.1 & 115.8 & 59.1 & 32.1 \\
    \cline{3-3}
     & & time& 22.1& 20.6& 20.2&39.6 \\
    \cline{2-7}
    &\multirow{2}{*}{\cite{huang2022cheetah}$^*$}& comm.& \textbf{14.1}& \textbf{9.9}& \textbf{10.8}& \textbf{17.2}\\
    \cline{3-3}
     & & time& \textbf{1.4}& \textbf{1.2}& \textbf{1.2}& \textbf{1.77}\\
    \cline{2-7}
    &\multirow{2}{*}{\cite{qiaosp24}}& comm.& 295.9& 154.7& 102.4& 99.9\\
    \cline{3-3}
     & & time& 15.5&11.8 & 11.1& 11\\
    \cline{2-7}
    &\multirow{4}{*}{\rotatebox{90}{\makebox[0pt][c]{\textbf{PrivQJ}}}}& \multirow{2}{*}{comm.}& \multirow{2}{*}{299.4}& \multirow{2}{*}{159.3}&  \multirow{2}{*}{108.3}& \multirow{2}{*}{104.7}\\
     & & & & & & \\
     \cline{3-3}
     & & \multirow{2}{*}{time}& \multirow{2}{*}{15.3}& \multirow{2}{*}{11.7}& \multirow{2}{*}{11.2}& \multirow{2}{*}{11}\\
      & & & & & & \\     
    \hline
    \multirow{6}{*}{\rotatebox{90}{\makebox[0pt][c]{WAN$_1$}}}& \multirow{1}{*}{\cite{rathee2020cryptflow2}}& time& 41.7& 30.7& 25.5& 42.9\\
    \cline{2-7}
    &\multirow{1}{*}{\cite{huang2022cheetah}$^*$}& time& \textbf{2.9}& \textbf{2.4}& \textbf{2.5}& \textbf{3.7}\\
    \cline{2-7}
    &\multirow{1}{*}{\cite{qiaosp24}}& time& 40.7& 25.3&20.5 & 20.1\\
    \cline{2-7}
    &\multirow{3}{*}{\rotatebox{90}{\makebox[0pt][c]{\textbf{PrivQJ}}}}& \multirow{3}{*}{time}& \multirow{3}{*}{40.5}& \multirow{3}{*}{25.5}& \multirow{3}{*}{20.7}& \multirow{3}{*}{20.3}\\
     & & & & & & \\
      & & & & & & \\     
    \hline
    \multirow{6}{*}{\rotatebox{90}{\makebox[0pt][c]{WAN$_2$}}}& \multirow{1}{*}{\cite{rathee2020cryptflow2}}& time& 44.7& 32.9& 27.3& 44.1\\
    \cline{2-7}
    &\multirow{1}{*}{\cite{huang2022cheetah}$^*$}& time& \textbf{3.9}& \textbf{2.9}& \textbf{3}&\textbf{3.9} \\
    \cline{2-7}
    &\multirow{1}{*}{\cite{qiaosp24}}& time& 44.6& 28.1& 22.3&21.8 \\
    \cline{2-7}
    &\multirow{3}{*}{\rotatebox{90}{\makebox[0pt][c]{\textbf{PrivQJ}}}}& \multirow{3}{*}{time}&\multirow{3}{*}{43.6} &\multirow{3}{*}{27.3} &\multirow{3}{*}{22.2} &\multirow{3}{*}{21.8} \\
     & & & & & & \\
      & & & & & & \\     
    \hline
    \multirow{6}{*}{\rotatebox{90}{\makebox[0pt][c]{WAN$_3$}}}& \multirow{1}{*}{\cite{rathee2020cryptflow2}}& time& 32.8& 26.6& 23.5& 41.8\\
    \cline{2-7}
    &\multirow{1}{*}{\cite{huang2022cheetah}$^*$}& time& \textbf{3}& \textbf{2.3}&\textbf{2.1} &\textbf{2.9} \\
    \cline{2-7}
    &\multirow{1}{*}{\cite{qiaosp24}}& time& 29.2&19.7 &16.4 &16.3 \\
    \cline{2-7}
    &\multirow{3}{*}{\rotatebox{90}{\makebox[0pt][c]{\textbf{PrivQJ}}}}& \multirow{3}{*}{time}& \multirow{3}{*}{28.5}& \multirow{3}{*}{19}& \multirow{3}{*}{16.3}&\multirow{3}{*}{16} \\
     & & & & & & \\
      & & & & & & \\     
    \hline
    \multirow{6}{*}{\rotatebox{90}{\makebox[0pt][c]{WAN$_4$}}}& \multirow{1}{*}{\cite{rathee2020cryptflow2}}& time& 35.8&28.9 &25 &43.1 \\
    \cline{2-7}
    &\multirow{1}{*}{\cite{huang2022cheetah}$^*$}& time& \textbf{2.8}&\textbf{2.6} &\textbf{2.6} &\textbf{3.4} \\
    \cline{2-7}
    &\multirow{1}{*}{\cite{qiaosp24}}& time& 32.9& 21.8& 18.4& 18\\
    \cline{2-7}
    &\multirow{3}{*}{\rotatebox{90}{\makebox[0pt][c]{\textbf{PrivQJ}}}}& \multirow{3}{*}{time}& \multirow{3}{*}{31.7}& \multirow{3}{*}{20.7}& \multirow{3}{*}{17.8}&\multirow{3}{*}{17.5} \\
     & & & & & & \\
      & & & & & & \\     
    \hline
    \hline
    \multicolumn{7}{c}{\textbf{(b)} \textit{Online cost}, communication cost (comm.) \textit{in MiB} and time cost \textit{in second}}\\
    \hline
    \multirow{10}{*}{\rotatebox{90}{\makebox[0pt][c]{LAN}}}& \multirow{2}{*}{\cite{rathee2020cryptflow2}}& comm.& 228.1 & 115.8 & 59.1 & 32.1 \\
    \cline{3-3}
     & & time& 22.1& 20.6& 20.2&39.6 \\
    \cline{2-7}
    &\multirow{2}{*}{\cite{huang2022cheetah}$^*$}& comm.& \textbf{14.1}& \textbf{9.9}& \textbf{10.8}& \textbf{17.2}\\
    \cline{3-3}
     & & time& \textbf{1.4}& 1.2& 1.2& 1.77\\
    \cline{2-7}
    &\multirow{2}{*}{\cite{qiaosp24}}& comm.& 203.4& 103.2& 52.6& 26.9\\
    \cline{3-3}
     & & time& 1.5&0.8 & \textbf{0.4}& \textbf{0.2}\\
    \cline{2-7}
    &\multirow{4}{*}{\rotatebox{90}{\makebox[0pt][c]{\textbf{PrivQJ}}}}& \multirow{2}{*}{comm.}& \multirow{2}{*}{206.9}& \multirow{2}{*}{107.8}&  \multirow{2}{*}{58.6}& \multirow{2}{*}{31.7}\\
     & & & & & & \\
     \cline{3-3}
     & & \multirow{2}{*}{time}& \multirow{2}{*}{\textbf{1.4}}& \multirow{2}{*}{\textbf{0.7}}& \multirow{2}{*}{\textbf{0.4}}& \multirow{2}{*}{\textbf{0.2}}\\
      & & & & & & \\     
    \hline
    \multirow{6}{*}{\rotatebox{90}{\makebox[0pt][c]{WAN$_1$}}}& \multirow{1}{*}{\cite{rathee2020cryptflow2}}& time& 41.7& 30.7& 25.5& 42.9\\
    \cline{2-7}
    &\multirow{1}{*}{\cite{huang2022cheetah}$^*$}& time& \textbf{2.9}& \textbf{2.4}& \textbf{2.5}& 3.7\\
    \cline{2-7}
    &\multirow{1}{*}{\cite{qiaosp24}}& time& 19.1& 10.1&5.5 & \textbf{3.1}\\
    \cline{2-7}
    &\multirow{3}{*}{\rotatebox{90}{\makebox[0pt][c]{\textbf{PrivQJ}}}}& \multirow{3}{*}{time}& \multirow{3}{*}{18.9}& \multirow{3}{*}{10}& \multirow{3}{*}{5.7}& \multirow{3}{*}{3.3}\\
     & & & & & & \\
      & & & & & & \\     
    \hline
    \multirow{6}{*}{\rotatebox{90}{\makebox[0pt][c]{WAN$_2$}}}& \multirow{1}{*}{\cite{rathee2020cryptflow2}}& time& 44.7& 32.9& 27.3& 44.1\\
    \cline{2-7}
    &\multirow{1}{*}{\cite{huang2022cheetah}$^*$}& time& \textbf{3.9}& \textbf{2.9}& \textbf{3}&\textbf{3.9} \\
    \cline{2-7}
    &\multirow{1}{*}{\cite{qiaosp24}}& time& 21.9& 12.1& 6.8&4.1 \\
    \cline{2-7}
    &\multirow{3}{*}{\rotatebox{90}{\makebox[0pt][c]{\textbf{PrivQJ}}}}& \multirow{3}{*}{time}&\multirow{3}{*}{21.2} &\multirow{3}{*}{11.5} &\multirow{3}{*}{6.7} &\multirow{3}{*}{4.1} \\
     & & & & & & \\
      & & & & & & \\     
    \hline
    \multirow{6}{*}{\rotatebox{90}{\makebox[0pt][c]{WAN$_3$}}}& \multirow{1}{*}{\cite{rathee2020cryptflow2}}& time& 32.8& 26.6& 23.5& 41.8\\
    \cline{2-7}
    &\multirow{1}{*}{\cite{huang2022cheetah}$^*$}& time& \textbf{3}& \textbf{2.3}&\textbf{2.1} &2.9 \\
    \cline{2-7}
    &\multirow{1}{*}{\cite{qiaosp24}}& time& 11&6.2 &3.5 &2.2 \\
    \cline{2-7}
    &\multirow{3}{*}{\rotatebox{90}{\makebox[0pt][c]{\textbf{PrivQJ}}}}& \multirow{3}{*}{time}& \multirow{3}{*}{10.6}& \multirow{3}{*}{5.6}& \multirow{3}{*}{3.3}&\multirow{3}{*}{\textbf{1.9}} \\
     & & & & & & \\
      & & & & & & \\     
    \hline
    \multirow{6}{*}{\rotatebox{90}{\makebox[0pt][c]{WAN$_4$}}}& \multirow{1}{*}{\cite{rathee2020cryptflow2}}& time& 35.8&28.9 &25 &43.1 \\
    \cline{2-7}
    &\multirow{1}{*}{\cite{huang2022cheetah}$^*$}& time& \textbf{2.8}&\textbf{2.6} &\textbf{2.6} &3.4 \\
    \cline{2-7}
    &\multirow{1}{*}{\cite{qiaosp24}}& time& 13.9& 8& 4.8& 3.2\\
    \cline{2-7}
    &\multirow{3}{*}{\rotatebox{90}{\makebox[0pt][c]{\textbf{PrivQJ}}}}& \multirow{3}{*}{time}& \multirow{3}{*}{12.9}& \multirow{3}{*}{7}& \multirow{3}{*}{4.3}&\multirow{3}{*}{\textbf{2.6}} \\
     & & & & & & \\
      & & & & & & \\     
    \hline    
    \multicolumn{7}{l}{$^*$ with more efficient HE and OT backends.}\\
    \end{tabular}
\end{table*}

\subsection{Efficiency When Plugged In Neural Models}
We further evaluate the performance of PrivQJ in neural model and Table \ref{eval::modeladded} shows the concrete comparison when applying different system for the queue jumping. Concretely, for neural models such as AlexNet and ResNet which left idle slots in most of the ciphertext, the added waiting cost is very small. Meanwhile, the cost increases when there is no idle slot in the ciphertext as a new set of ciphertext for prior input is needed. For example, there is no idle slot in the server-encrypted ciphertext in some blocks of VGG, i.e., the ciphertext for $\bm{h}_1$ and $\{\bm{x}_1\times(\bm{1}-\bm{2}\times\bm{h}_1)\}$ in Figure \ref{fig:reluconv}, which forces PrivQJ to invoke a totally new run at online and the time becomes longer when the bandwidth decreases. On the other hand, most computation sizes in neural models always left slots in the ciphertext for $\widehat{\bm{r}_0}$, which helps to mitigate a large part of computation for prior input because the major time cost comes from the HE computation among ciphertext for $\widehat{\bm{r}_0}$ and it makes PrivQJ remain much more efficient regarding the overall cost.

We then investigate the performance of in-queue input in PrivQJ to make sure the efficiency of original input remains advantageous. As shown in Table \ref{eval::modelinqueue}, PrivQJ always has better time overhead for in-queue inputs and competitive communication cost compared with FIT.

\begin{table}
    \centering
    \footnotesize
    \caption{Added cost to in-queue input for neural models.
}\label{eval::modeladded}
\resizebox{\columnwidth}{!}{%
    \begin{tabular}{c|ccccc}
    \hline
    \multicolumn{6}{c}{\textbf{(a)} \textit{Overall cost}, communication cost (comm.) \textit{in MiB} and time cost \textit{in second}}\\
    \hline
    \multicolumn{3}{c}{Neural Models}& AlexNet \cite{krizhevsky2012imagenet} & VGG \cite{simonyan2014very} & ResNet \cite{he2016deep}\\
    \hline
    \multirow{12}{*}{\rotatebox{90}{\makebox[0pt][c]{LAN}}}& \multirow{4}{*}{\cite{rathee2020cryptflow2}}& comm.& 659 & 7567 & 1740\\
     & & speedup& \textbf{260}$\times$& \textbf{2.9}$\times$& \textbf{153}$\times$\\
    \cline{3-3}
     & & time& 189& 1326& 410\\
     & & speedup& \textbf{439}$\times$& \textbf{51}$\times$& \textbf{410}$\times$\\
    \cline{2-6}
    &\multirow{4}{*}{\cite{qiaosp24}}& comm.& 528& 4318& 2611\\
     & & speedup& \textbf{208}$\times$& \textbf{1.6}$\times$& \textbf{230}$\times$\\
    \cline{3-3}
     & & time& 89& 796& 197\\
     & & speedup& \textbf{207}$\times$& \textbf{31}$\times$& \textbf{197}$\times$\\
    \cline{2-6}
    &\multirow{4}{*}{\rotatebox{90}{\makebox[0pt][c]{\textbf{PrivQJ}}}}& \multirow{2}{*}{comm.}& \multirow{2}{*}{2.5}& \multirow{2}{*}{2557}&  \multirow{2}{*}{11.3}\\
     & & & & &\\
     \cline{3-3}
     & & \multirow{2}{*}{time}& \multirow{2}{*}{0.4}& \multirow{2}{*}{25.6}& \multirow{2}{*}{1}\\
      & & & & &\\     
    \hline
    \multirow{7}{*}{\rotatebox{90}{\makebox[0pt][c]{WAN$_1$}}}& \multirow{2}{*}{\cite{rathee2020cryptflow2}}& time& 246& 1953& 563\\
     & & speedup& \textbf{569}$\times$& \textbf{8.1}$\times$& \textbf{563}$\times$\\
    \cline{2-6}
    &\multirow{2}{*}{\cite{qiaosp24}}& time& 136& 1160&426\\
     & & speedup& \textbf{315}$\times$& \textbf{4.8}$\times$& \textbf{426}$\times$\\
    \cline{2-6}
    &\multirow{3}{*}{\rotatebox{90}{\makebox[0pt][c]{\textbf{PrivQJ}}}}& \multirow{3}{*}{time}& \multirow{3}{*}{0.4}& \multirow{3}{*}{238}& \multirow{3}{*}{1}\\
     & & & & &\\
      & & & & &\\     
    \hline
    \hline
    \multicolumn{6}{c}{\textbf{(b)} \textit{Online cost}, communication cost (comm.) \textit{in MiB} and time cost \textit{in second}}\\
    \hline
    \multirow{12}{*}{\rotatebox{90}{\makebox[0pt][c]{LAN}}}& \multirow{4}{*}{\cite{rathee2020cryptflow2}}& comm.& 659 &  7567& 1740\\
     & & speedup& \textbf{260}$\times$& \textbf{3.1}$\times$& \textbf{153}$\times$\\
    \cline{3-3}
     & & time& 189& 1326& 410\\
     & & speedup& \textbf{439}$\times$& \textbf{54}$\times$& \textbf{410}$\times$\\
    \cline{2-6}
    &\multirow{4}{*}{\cite{qiaosp24}}& comm.& 249& 2844& 1544\\
     & & speedup& \textbf{98}$\times$& \textbf{1.18}$\times$& \textbf{136}$\times$\\
    \cline{3-3}
     & & time& 2.3& 28.4& 11.6\\
     & & speedup& \textbf{5.2}$\times$& \textbf{1.17}$\times$& \textbf{11}$\times$\\
    \cline{2-6}
    &\multirow{4}{*}{\rotatebox{90}{\makebox[0pt][c]{\textbf{PrivQJ}}}}& \multirow{2}{*}{comm.}& \multirow{2}{*}{2.5}& \multirow{2}{*}{2410}&  \multirow{2}{*}{11.3}\\
     & & & & &\\
     \cline{3-3}
     & & \multirow{2}{*}{time}& \multirow{2}{*}{0.4}& \multirow{2}{*}{24}& \multirow{2}{*}{1}\\
      & & & & &\\     
    \hline
    \multirow{7}{*}{\rotatebox{90}{\makebox[0pt][c]{WAN$_1$}}}& \multirow{2}{*}{\cite{rathee2020cryptflow2}}& time& 246& 1954& 563\\
     & & speedup& \textbf{569}$\times$& \textbf{8.7}$\times$& \textbf{563}$\times$\\
    \cline{2-6}
    &\multirow{2}{*}{\cite{qiaosp24}}& time& 25.7& 268&151\\
     & & speedup& \textbf{59}$\times$& \textbf{1.19}$\times$& \textbf{151}$\times$\\
    \cline{2-6}
    &\multirow{3}{*}{\rotatebox{90}{\makebox[0pt][c]{\textbf{PrivQJ}}}}& \multirow{3}{*}{time}& \multirow{3}{*}{0.4}& \multirow{3}{*}{224}& \multirow{3}{*}{1}\\
     & & & & &\\
      & & & & &\\     
    \hline
    \end{tabular}
    }
\end{table}

\begin{table}
    \centering
    \footnotesize
    \caption{Efficiency of in-queue input for neural models.
}\label{eval::modelinqueue}
\resizebox{\columnwidth}{!}{%
    \begin{tabular}{c|ccccc}
    \hline
    \multicolumn{6}{c}{\textbf{(a)} \textit{Overall cost}, communication cost (comm.) \textit{in MiB} and time cost \textit{in second}}\\
    \hline
    \multicolumn{3}{c}{Neural Models}& AlexNet \cite{krizhevsky2012imagenet} & VGG \cite{simonyan2014very} & ResNet \cite{he2016deep}\\
    \hline
    \multirow{8}{*}{\rotatebox{90}{\makebox[0pt][c]{LAN}}}& \multirow{2}{*}{\cite{rathee2020cryptflow2}}& comm.& 659 & 7567 & 1740\\
    \cline{3-3}
     & & time& 189& 1326& 410\\
    \cline{2-6}
    &\multirow{2}{*}{\cite{qiaosp24}}& comm.& 528& \textbf{4318}& \textbf{2611}\\
    \cline{3-3}
     & & time& 89& 796& 197\\
    \cline{2-6}
    &\multirow{4}{*}{\rotatebox{90}{\makebox[0pt][c]{\textbf{PrivQJ}}}}& \multirow{2}{*}{comm.}& \multirow{2}{*}{\textbf{526}}& \multirow{2}{*}{4331}&  \multirow{2}{*}{2686}\\
     & & & & &\\
     \cline{3-3}
     & & \multirow{2}{*}{time}& \multirow{2}{*}{\textbf{83.3}}& \multirow{2}{*}{\textbf{730.6}}& \multirow{2}{*}{\textbf{196.8}}\\
      & & & & &\\     
    \hline
    \multirow{5}{*}{\rotatebox{90}{\makebox[0pt][c]{WAN$_1$}}}& \multirow{1}{*}{\cite{rathee2020cryptflow2}}& time& 246& 1953& 563\\
    \cline{2-6}
    &\multirow{1}{*}{\cite{qiaosp24}}& time& 136& 1160&\textbf{426}\\
    \cline{2-6}
    &\multirow{3}{*}{\rotatebox{90}{\makebox[0pt][c]{\textbf{PrivQJ}}}}& \multirow{3}{*}{time}& \multirow{3}{*}{\textbf{128}}& \multirow{3}{*}{\textbf{1092}}& \multirow{3}{*}{428}\\
     & & & & &\\
      & & & & &\\     
    \hline
    \hline
    \multicolumn{6}{c}{\textbf{(b)} \textit{Online cost}, communication cost (comm.) \textit{in MiB} and time cost \textit{in second}}\\
    \hline
    \multirow{8}{*}{\rotatebox{90}{\makebox[0pt][c]{LAN}}}& \multirow{2}{*}{\cite{rathee2020cryptflow2}}& comm.& 659 &  7567& 1740\\
    \cline{3-3}
     & & time& 189& 1326& 410\\
    \cline{2-6}
    &\multirow{2}{*}{\cite{qiaosp24}}& comm.& \textbf{249}& \textbf{2844}& \textbf{1544}\\
    \cline{3-3}
     & & time& 2.3& 28.4& 11.6\\
    \cline{2-6}
    &\multirow{4}{*}{\rotatebox{90}{\makebox[0pt][c]{\textbf{PrivQJ}}}}& \multirow{2}{*}{comm.}& \multirow{2}{*}{256}& \multirow{2}{*}{2857}&  \multirow{2}{*}{1620}\\
     & & & & &\\
     \cline{3-3}
     & & \multirow{2}{*}{time}& \multirow{2}{*}{\textbf{2.1}}& \multirow{2}{*}{\textbf{27}}& \multirow{2}{*}{\textbf{10.8}}\\
      & & & & &\\     
    \hline
    \multirow{5}{*}{\rotatebox{90}{\makebox[0pt][c]{WAN$_1$}}}& \multirow{1}{*}{\cite{rathee2020cryptflow2}}& time& 246& 1954& 563\\
    \cline{2-6}
    &\multirow{1}{*}{\cite{qiaosp24}}& time& 25.7& 268&\textbf{151}\\
    \cline{2-6}
    &\multirow{3}{*}{\rotatebox{90}{\makebox[0pt][c]{\textbf{PrivQJ}}}}& \multirow{3}{*}{time}& \multirow{3}{*}{\textbf{24.7}}& \multirow{3}{*}{\textbf{262.5}}& \multirow{3}{*}{151.6}\\
     & & & & &\\
      & & & & &\\     
    \hline
    \end{tabular}
    }
\end{table}

Finally, we analyze the needed batch size for the prior input to have a more comprehensive exploration to PrivQJ, and Table \ref{tab:bsize} demonstrates the distribution in major blocks. Specifically, the batch size (b-size) indicates the number of in-queue inputs in a batch to achieve almost-free computation for prior input, and the b-size for online is the number of in-queue inputs for a full slot recycling in cipher of $\textit{\textbf{t}}$ as described in Section \ref{slot:recycle} while the b-size for offline is the number of in-queue inputs for a full slot recycling in cipher of $\widehat{\bm{r}_0}$ as described in Section \ref{r0kslotrec}. As shown in Table \ref{tab:bsize}, the maximum batch size is moderate for online computation, e.g., 102, and b-size $=1$ indicates no idle slots in associated cipher and a new run is needed at online. Meanwhile, the offline needs more in-queue inputs which is generally tolerable since it is input-independent. Moreover, appropriate division of the input helps to adjust the sizes to have smaller b-sizes at both offline and online.

\begin{table}
    \centering
    \caption{Distribution of batch size ({b-size}) in major blocks.}
    \label{tab:bsize}
    \begin{tabular}{c|c|c}
    \hline
    $H_i, C_i, f_h, C_o$& b-size at {online} & b-size at offline \\
    \hline
    56,64,3,64& 49 & 4\\
    28,128,3,128 & 17 & 30\\
    14,256,3,256 & 7 & 82\\
    7,512,3,512&4&990\\
    112,64,3,128&1&4\\
    56,128,3,256&1&4\\
    56,256,3,256&1&4\\
    28,256,3,512&49&30\\
    28,512,3,512&1&30\\
    14,512,3,512&17&82\\
    27,96,5,256&19&130\\
    13,256,3,384&8&144\\
    13,384,3,384&102&144\\
    13,384,3,256&102&144\\
    \hline
    \end{tabular}
\end{table}

\section{Related Works}\label{related:work}
PP-MLaaS has attracted significant attention in recent decade, with solutions generally falling into two categories. The first class follows an \textit{offline-online paradigm}, where computationally expensive cryptographic operations are performed in a pre-processing phase before the client submits inputs, and lightweight operations dominate during online inference, leading to significant latency reduction. Representative protocols, such as MiniONN \cite{liu2017oblivious}, DELPHI \cite{mishra2020delphi} and FIT \cite{qiaosp24}, combine HE with MPC techniques to allow much lower online cost. 

The second class avoids pre-processing and instead operates in a \textit{purely online setting}, where all cryptographic computation is executed at query time. While it removes the need for cost-intensive offline, the computation and communication cost at inference remains relatively high. CryptoNets \cite{cryptonets} is the pioneering work that shows the feasibility of HE in PP-MLaaS. After that, a series of landmark systems have been proposed for better efficiency, such as HE-MPC-based frameworks SecureML \cite{mohassel2017secureml}, GAZELLE \cite{juvekar2018gazelle}, CrypTFlow2 \cite{rathee2020cryptflow2} and Cheetah \cite{huang2022cheetah}, HE-based frameworks E2DM \cite{jiang2018secure} and LOHEN \cite{nam2025lohen}, Garbled Circuit (GC)-based frameworks DeepSecure \cite{rouhani2018deepsecure} and XONN \cite{riazi2019xonn}, and their variants such as SIRNN \cite{rathee2021sirnn} and SecFloat \cite{DBLP:conf/sp/RatheeB00CR22}.
Our work \textrm{PrivQJ} lies in the offline-online category, with much lower cost when the queue jumping happens. 

A further distinction arises between \textit{2-party} and \textit{3-party} privacy-preserving inference. In 3-party protocols, an additional non-colluding server is introduced to assist in secret sharing, which helps to reduce the overall cryptographic overhead. While these approaches, such as ABY3 \cite{mohassel2018aby3}, ABY2.0 \cite{patra2021aby2} and SHAFT \cite{kei2025shaft}, demonstrate excellent efficiency gain, they require stronger trust for more entities. By contrast, our work operates in a 2-party setting between the client and the server, without reliance on third party. This makes it easier to deploy in real-world applications where involving additional service providers may be impractical or undesirable, while still enjoys competitive efficiency benefits.

Finally, \textrm{PrivQJ} targets on \textit{private inference for CNNs} while the community has devoted increasing attention to \textit{privacy-preserving inference for large language models (LLMs)}, which addresses efficiency challenges with much-larger model sizes. While these approaches, such as BOLT \cite{pangsp24}, BumbleBee \cite{cryptoeprint:2023/1678}, and NEXUS \cite{cryptoeprint:2024/136}, are important for the growing field of natural language processing, there is no study about privacy-preserving queue jumping in both CNNs and LLMs models and they all need non-trivial efforts to investigate associated problems. By focusing on CNNs in an offline-online scheme under a 2-party setting, our work provides a starting exploration for privacy-preserving queue jumping in private inference and exploring the problem in LLMs forms another line of works that are worth investigation.

\section{Conclusion}\label{conclusion}
In this paper, we have initiated the study of privacy-preserving queue jumping in batched inference and propose PrivQJ, a novel framework that enables efficient priority handling without degrading overall system performance. PrivQJ exploits shared computation across inputs via in-processing slot recycling, allowing prior inputs to be piggybacked onto ongoing batch computation with almost no additional cryptographic cost. Both theoretical analysis and experimental results have demonstrated noticeable reduction in overhead compared to state-of-the-art PP-MLaaS systems. 
While \textrm{PrivQJ} deals with convolutional neural networks in this paper, adapting it to other types of models forms our next-step direction.



\section*{Acknowledgment}
The abstract is first generated using \href{https://chatgpt.com/}{ChatGPT} and is then manually verified and tailored for correct and concise description.


\bibliographystyle{IEEEtran}
\bibliography{2025-PriorClient-IEEE}

\vfill

\end{document}


\begin{appendices}
\begin{figure}[!b]
\centering
\includegraphics[trim={0.1cm 4.5cm 12.1cm 0.1cm}, clip, scale=0.49]{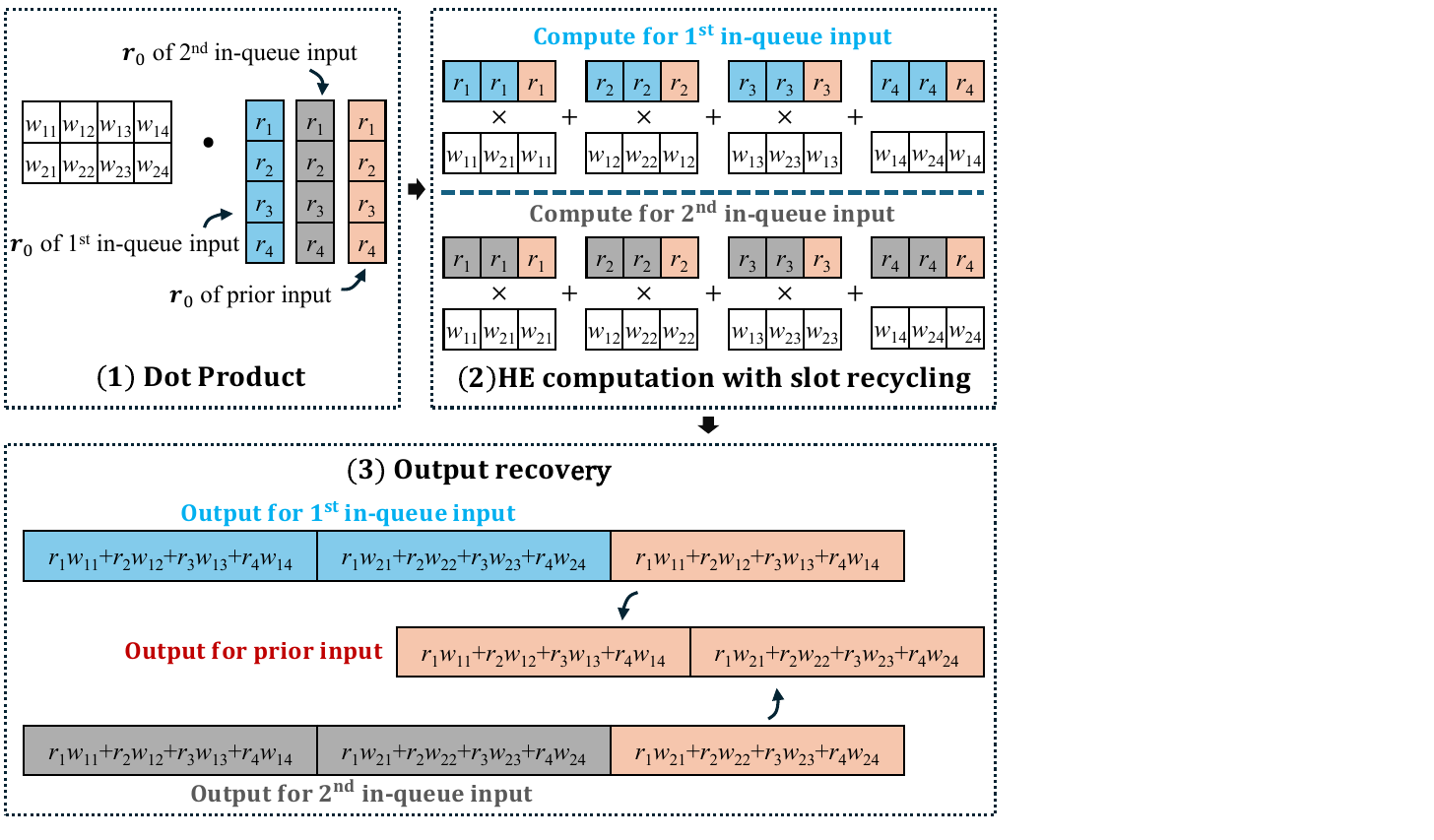}
\caption{HE computation for dot product with slot recycling.}
\label{fig:fc}
\end{figure}

\section{Deal with Other Neural Functions}\label{appdx:1}

While we apply model adjustment to other functions similar with FIT framework to make full use of proposed module, we would like to mention three points. 
First, the dot product in neural models is treated similarly with convolution and Figure~\ref{fig:fc} gives a concrete example to piggyback the computation for prior input within the process for in-queue inputs.
Second, the cost to compute other functions for prior inputs is evenly shared among the computation for associated in-queue inputs in a batch and we evaluate through experiments that the extra cost amortized to each in-queue input is negligible, and we refer to Section~\ref{eval} for more discussion.
Finally, the number of needed in-queue inputs for a prior input could be varied when computing $f_{\mathrm c}f_{\mathrm r}(\bm{x})$ with different sizes of $\bm{x}$ and/or $\bm{k}$ in a neural model, and we could make segmentation over input to produce smaller numbers.

\section{Security Proof}\label{appdx}
\textrm{Theorem} 1. \textit{The offline computation for in-queue inputs, as shown in  Figure~\ref{fig:reluconv}, is secure in the presence of semi-honest adversaries, if the HE Encryption is semantically secure.}

\textit{Proof.} In the real world, $\mathcal{S}$ and $\mathcal{C}$ follow the protocol while both parties in the ideal world have access to a trust third party TTP which returns \doublebox{$\widehat{\bm{r}_0}$}$\ast\bm{k}$ to $\mathcal{S}$, and \fbox{$\bm{h}_1$} and \fbox{$\bm{x}_1\times(\bm{1}-\bm{2}\times\bm{h}_1)$} to $\mathcal{C}$. The execution in both worlds are coordinated by environment $\bm{\varepsilon}$ which chooses inputs to $\mathcal{S}$ and $\mathcal{C}$, and distinguishes between real world and ideal world. We aim to demonstrate that the view of adversary in real world is indistinguishable to that in ideal world.

\textit{Security against a semi-honest server.} We prove security against a semi-honest server by constructing an ideal-world simulator $sim$ that performs the following:

(1) receives $\bm{k}$, $\bm{x}_1$ and $\bm{h}_1$ from $\bm{\varepsilon}$, $sim$ sends them to TTP and receives \doublebox{$\widehat{\bm{r}_0}$}$\ast\bm{k}$;

(2) stars running $\mathcal{S}$ on inputs $\bm{k}$, $\bm{x}_1$ and $\bm{h}_1$;

(3) encrypts the crafted data ${\bm{r}_0^{sim}}=\{\bm{0}\}^{H_fW_fC_i\times H_oW_o}$ with a randomly-selected public key, and sends that new ciphertext, denoted as \raisebox{0.5ex}{\tikz[baseline]{\node[draw, dashed, inner sep=2pt] {{${\bm{r}_0^{sim}}$}};}}, to $\mathcal{S}$;

(4) outputs whatever $\mathcal{S}$ outputs.

The view of $\mathcal{S}$ in real execution is \doublebox{$\widehat{\bm{r}_0}$} while its view in ideal world is \raisebox{0.5ex}{\tikz[baseline]{\node[draw, dashed, inner sep=2pt] {{${\bm{r}_0^{sim}}$}};}}. Since \doublebox{$\widehat{\bm{r}_0}$} is computationally indistinguishable from \raisebox{0.5ex}{\tikz[baseline]{\node[draw, dashed, inner sep=2pt] {{${\bm{r}_0^{sim}}$}};}} based on the semantic security of HE encryption, the output distribution of $\bm{\varepsilon}$ in real world is computationally indistinguishable from that in ideal world.

\textit{Security against a semi-honest client.} We prove security against a semi-honest client by constructing an ideal-world simulator $sim$ that performs the following:

(1) receives $\bm{r}_0$ from $\bm{\varepsilon}$, $sim$ sends it to TTP and receives \fbox{$\bm{h}_1$} and \fbox{$\bm{x}_1\times(\bm{1}-\bm{2}\times\bm{h}_1)$};

(2) stars running $\mathcal{C}$ on input $\bm{r}_0$;

(3) encrypts the crafted data ${\bm{h}_1^{sim}}=\{\bm{0}\}^{C_i\times H_i\times W_i}$ with a randomly-selected public key, and sends two copies of that new ciphertext, denoted as $\{$\raisebox{0.5ex}{\tikz[baseline]{\node[draw, dashed, inner sep=2pt] {$\bm{h}_1^{sim}$};}}$\}^2$, to $\mathcal{C}$;

(4) outputs whatever $\mathcal{C}$ outputs.

The views of $\mathcal{C}$ in real execution are two ciphertext namely \fbox{$\bm{h}_1$} and \fbox{$\bm{x}_1\times(\bm{1}-\bm{2}\times\bm{h}_1)$} while its views in ideal world are $\{$\raisebox{0.5ex}{\tikz[baseline]{\node[draw, dashed, inner sep=2pt] {$\bm{h}_1^{sim}$};}}$\}^2$. Since \fbox{$\bm{h}_1$} and \fbox{$\bm{x}_1\times(\bm{1}-\bm{2}\times\bm{h}_1)$} are computationally indistinguishable from $\{$\raisebox{0.5ex}{\tikz[baseline]{\node[draw, dashed, inner sep=2pt] {$\bm{h}_1^{sim}$};}}$\}^2$ based on the semantic security of HE encryption, the output distribution of $\bm{\varepsilon}$ in real world is computationally indistinguishable from that in ideal world.

\textrm{Theorem} 2. \textit{The online computation for in-queue inputs, as shown in  Figure~\ref{fig:reluconv}, is secure in the presence of semi-honest adversaries, if the OT and HE Encryption are semantically secure.}

\textit{Proof.} In the real world, $\mathcal{S}$ and $\mathcal{C}$ follow the protocol while both parties in the ideal world have access to a trust third party TTP which returns $\sum((\bm{x}_1\times\bm{h}_1+\bm{t})\ast\bm{k}+\widehat{\bm{r}_0}\ast\bm{k}-\widehat{\bm{x}_1})$ to $\mathcal{C}$. The execution in both worlds are coordinated by environment $\bm{\varepsilon}$ which chooses inputs to $\mathcal{S}$ and $\mathcal{C}$, and distinguishes between real world and ideal world. We aim to demonstrate that the view of adversary in real world is indistinguishable to that in ideal world.

\textit{Security against a semi-honest server.} We prove security against a semi-honest server by constructing an ideal-world simulator $sim$ that performs the following:

(1) receives $\bm{x}_1$, $\bm{h}_1$ and $\widehat{\bm{x}_1}$ from $\bm{\varepsilon}$, $sim$ sends them to TTP;

(2) stars running $\mathcal{S}$ on inputs $\bm{x}_1$, $\bm{h}_1$ and $\widehat{\bm{x}_1}$, and receives $\widehat{\bm{h}_0}\in\mathbb{Z}_2^{C_i\times H_i\times W_i}$;

(3) encrypts the crafted data ${\bm{t}^{sim}}$ $\scriptstyle\overset{\$}{\gets}$ $\mathbb{Z}_p^{C_i\times H_i\times W_i+\widehat{s}}$ with $\mathcal{S}$'s public key, and sends that ciphertext, denoted as \fbox{${\bm{t}^{sim}}$}, to $\mathcal{S}$;

(4) outputs whatever $\mathcal{S}$ outputs.

On the one hand, the semantic security of OT guarantees the computational indistinguishability of messages received at $\mathcal{S}$ for computing $f'_{\mathrm r}(\bm{x})$ in real and ideal executions. On the other hand, the view of $\mathcal{S}$ after computing $f'_{\mathrm r}(\bm{x})$ is $\bm{t}$ in real world while its view in ideal world is ${\bm{t}^{sim}}$. Thus we only need to show that any element in $\bm{t}$ is indistinguishable from a random number in ${\bm{t}^{sim}}$. It is clearly true since ${\bm{r}_{0}}$ in $\bm{t}$ is randomly chosen.
Therefore, the output distribution of $\bm{\varepsilon}$ in real world is computationally indistinguishable from that in ideal world.

\textit{Security against a semi-honest client.} We prove security against a semi-honest client by constructing an ideal-world simulator $sim$ that performs the following:

(1) receives $\bm{x}_0$ and $\bm{r}_0$ from $\bm{\varepsilon}$, $sim$ sends it to TTP and receives $\sum((\bm{x}_1\times\bm{h}_1+\bm{t})\ast\bm{k}+\widehat{\bm{r}_0}\ast\bm{k}-\widehat{\bm{x}_1})$;

(2) stars running $\mathcal{C}$ on input $\bm{x}_0$ and $\bm{r}_0$, and receives \fbox{${\widehat{\bm{t}}}$} where $\widehat{\bm{t}}\in\mathbb{Z}_p^{C_i\times H_i\times W_i+\widehat{s}}$;

(3) splits $\sum((\bm{x}_1\times\bm{h}_1+\bm{t})\ast\bm{k}+\widehat{\bm{r}_0}\ast\bm{k}-\widehat{\bm{x}_1})$ into another group of vectors $\widehat{\bm{x}_1^{sim}}\in\{\mathbb{Z}_p^{N}\}^{C_o}$ which have the same structure as $((\bm{x}_1\times\bm{h}_1+\bm{t})\ast\bm{k}+\widehat{\bm{r}_0}\ast\bm{k}-\widehat{\bm{x}_1})$;

(4) encrypts $\widehat{\bm{x}_1^{sim}}$ with $\mathcal{C}$'s public key and sends the new ciphertext, denoted as \doublebox{$\widehat{\bm{x}_1^{sim}}$}, to $\mathcal{C}$;

(5) outputs whatever $\mathcal{C}$ outputs.

On the one hand, the semantic security of OT guarantees the computational indistinguishability of messages received at $\mathcal{C}$ for computing $f'_{\mathrm r}(\bm{x})$ in real and ideal executions. On the other hand, the view of $\mathcal{C}$ after computing $f'_{\mathrm r}(\bm{x})$ is $((\bm{x}_1\times\bm{h}_1+\bm{t})\ast\bm{k}+\widehat{\bm{r}_0}\ast\bm{k}-\widehat{\bm{x}_1})$ in real world while its view in ideal world is $\widehat{\bm{x}_1^{sim}}$. Thus we only need to show that any element in $((\bm{x}_1\times\bm{h}_1+\bm{t})\ast\bm{k}+\widehat{\bm{r}_0}\ast\bm{k}-\widehat{\bm{x}_1})$ is indistinguishable from a random number in $\widehat{\bm{x}_1^{sim}}$. It is clearly true since $\widehat{\bm{x}_1}$ is randomly chosen. At the end of simulation, $\mathcal{C}$ outputs $\sum\widehat{\bm{x}_1^{sim}}$, which is the same as that in real execution.
Therefore, the output distribution of $\bm{\varepsilon}$ in real world is computationally indistinguishable from that in ideal world.

\textrm{Theorem} 3. \textit{The online computation for prior inputs, as shown in  Figure~\ref{fig:reluconv}, is secure in the presence of semi-honest adversaries if the Additive Secret Sharing (ASS) is semantically secure.}

\textit{Proof.} Since it is a sharing of output $(\bm{x}_1\times\bm{h}_1+\bm{t})\ast\bm{k}$, the security is guaranteed based on the security of ASS.

As such, we prove that~\textrm{PrivQJ} is secure under semi-honest assumption based on the above three theorems.
\end{appendices}

\vfill